\documentclass[twocolumn]{aastex631}

\usepackage{wasysym}

\begin{document}

\title{The LUNIS-AGN Catalog: Trends of emission line velocity dispersion and surface brightness within the circumnuclear regions of Seyfert galaxies}

\author[0009-0007-6992-2555]{Dan Delaney}
\affiliation{Department of Physics, University of Alaska Fairbanks, AK 99775, USA}
\affiliation{Department of Physics and Astronomy, University of Alaska Anchorage, AK 99508, USA}

\author{Cassidy Berger}
\affiliation{Department of Physics and Astronomy, University of Alaska Anchorage, AK 99508, USA}

\author[0000-0002-4457-5733]{Erin Hicks}
\affiliation{Department of Physics and Astronomy, University of Alaska Anchorage, AK 99508, USA}
\affiliation{Department of Physics, University of Alaska Fairbanks, AK 99775, USA}

\author[0000-0003-1014-043X]{Leonard Burtscher}
\affiliation{Leiden Observatory, PO Box 9513, 2300RA, Leiden, Netherlands}

\author[0000-0002-0001-3587]{David Rosario}
\affiliation{School of Mathematics, Statistics and Physics, Newcastle University, Newcastle upon Tyne NE1 7RU, UK}

\author[0000-0002-2713-0628]{Francisco M{\"u}ller-S\'anchez}
\affiliation{Department of Physics and Materials Science, The University of Memphis, 3720 Alumni Avenue, Memphis, TN 38152, USA}

\author[0000-0001-6919-1237]{Matthew Malkan}
\affiliation{Department of Physics and Astronomy, UCLA, Los Angeles, CA 90095-1547, USA}



\begin{abstract}

We present a catalog of Local Universe Near-Infrared Seyfert  (LUNIS) \textit{K-}band integral field unit (IFU) data of 88 nearby Active Galactic Nuclei (AGN), curated from SINFONI/VLT and OSIRIS/Keck archival datasets. This catalog includes both type 1 and 2 Seyfert AGN probed down to scales of tens of parsecs with z $<$ 0.02 and spanning over five orders of magnitude in L$_{14-195keV}$ AGN luminosity. As part of this catalog we make publicly available for all galaxies the processed datacubes, a central 200 pc integrated spectrum, and two-dimensional maps of flux, velocity, and velocity dispersion for H$_2$ 1-0 S(1) 2.1218 $\mu$m, [Si VI] 1.9641 $\mu$m, and Br-$\gamma$ 2.1655 $\mu$m. The morphology and geometry of [Si VI], a tracer of AGN outflows, are reported for the 66$\%$ of galaxies with extended emission. We utilize this large sample to probe the behavior of molecular and ionized gas, identifying trends in the properties of the circumnuclear gas (surface brightness and velocity dispersion) with fundamental AGN properties (obscuration and X-ray luminosity). While there is significant variation in circumnuclear gas characteristics across the sample, we find molecular hydrogen to be less centrally concentrated and exhibit lower velocity dispersion relative to ionized gas. In addition, we find elevated molecular hydrogen surface brightness and decreased [Si VI] velocity dispersion in obscured relative to unobscured AGN. The [Si VI] and Br-$\gamma$ emission scale with L$_{14-195keV}$ X-ray luminosity, which, along with the elevated velocity dispersion compared to the molecular gas, verifies an association with AGN outflow processes.

\end{abstract}



\section{Introduction} \label{sec:intro}

The immense radiation emanating from AGN is understood to be the result of mass accreting into the supermassive black hole (SMBH) nested at the galactic center. Due to the thermo-viscous processes that occur as material accretes toward the SMBH, AGN are among the most luminous objects in the universe with emission spanning a broad range of wavelengths with most of the light produced by the accretion disk ranging between 30 to 300 nm \citep{shakura1973black, rees1984black, hickox2018}. Despite the luminous nature of the accretion disk, the nuclear radiation from the AGN is often obscured from view. Obscuration may be the result of either interstellar medium, the galactic disk itself, or on a smaller scale, dust obscuration which may (or may not) resemble a toroidal structure around the SMBH and likely manifests as a result of turbulent material kinematics of the local environment \citep{wada2012radiation, wada2015obscuring, hickox2018}. The anatomy of an AGN is currently understood via its major components including the central SMBH engine, its powerful accretion disk, the small scale (generally sub-pc) Broad Line Region (BLR), a geometrically and optically thick obscuring dusty toroidal structure shrouding the AGN, and to greater extents the circumnuclear disk of dusty/gaseous material in orbit of the AGN. Following this Unified AGN model, Seyfert 1 and Seyfert 2 galaxies were proposed to be intrinsically similar objects, with the primary difference being the viewing angle or orientation of the dusty obscuring torus relative to the observer \citep{antonucci1993unified, urry1995unified}. In this case Seyfert 2 AGN are expected to have the torus structure obscuring the view of the small scale circumnuclear BLR clouds around the SMBH, hence the lack of observed broad emission lines. However, challenging this picture, recent studies have identified differences across Seyfert types that can not be explained by viewing angle (e.g. properties of the nuclear dusty torus structure \citep{almeida2011testing, rosario2018llama}. 

The radiative energy released by the accretion disk provides feedback from the growth of the SMBH to the rest of the galaxy, influencing the motions of material around the AGN \citep{wada2012radiation, almeida2017nuclear}. However, the processes and physics which result in material falling in and fueling the AGN and outflows from the galactic nucleus remain enigmatic. The circumnuclear region (r $<$ 500 pc) acts as a bridge between the AGN and the rest of the galaxy; a first layer of information transfer between the SMBH and the greater galactic dynamics. As such, it is crucial to probe the nature of the torus and circumnuclear region (r $<$ 500 pc) to understand AGN feeding and feedback mechanisms. 
Near infrared (NIR) observations provide insight into the motion of the warm molecular and ionized gas around the AGN, allowing us to probe how material flows into, out of, and around the SMBH. Due to their relatively low luminosity nature, Seyfert type galaxies nuclear emission do not fully dominate the spectra, thus these objects provide an excellent opportunity to glean insight as to the nature of material in the central region. As such, high resolution NIR spectroscopic data of the nuclear region of local Seyfert AGN is critical to resolve and study the nature of material around the SMBH and to identify and characterize the relevant AGN feeding and feedback mechanisms. Archival NIR datasets such as those presented in this catalog as well as observations in other wavelengths from archival and future datasets, such as the mid-infrared observations from the James Webb Space Telescope (JWST) and observations from the Extremely Large Telescope (ELT), will be necessary to guide physical models and understand these systems.   

This study builds upon a foundation of knowledge established in previous multiwavelength works aimed at understanding the relationship between AGN energetics and inflows and outflows.  Such efforts have utilized multi-wavelength data collected via HST, VLT, ALMA, Gemini, and now JWST to characterize the AGN circumnuclear material. While this work acts to enhance our understanding, other studies have also performed investigations using relatively large sample sets. Notably, \cite{burtscher2015obscuration} analyzed non-stellar emission in IFU data for a sample of 51 Seyfert AGN in the NIR constraining differences between Seyfert 1 and Seyfert 2 AGN, \cite{muller2018keck} also focused on continuum emission in the NIR and analyzed 40 Seyfert AGN highlighting a correlation between X-ray luminosity and \textit{K}-band luminosity. Additionally, \cite{garcia2021galaxy} investigated feedback in a sample of 19 Seyfert AGN finding evidence that imprint of AGN feedback on the host galaxy correlates with X-ray luminosity. \cite{ruschel2021agnifs}, \cite{riffel2023agnifs}, and \cite{schonell2019gemini} used Gemini IFU data to study molecular and ionized gas kinematics and distributions in nearby X-ray active galaxies (30, 33 and 6 AGN respectively) in the NIR. In addition, numerous multi-wavelength studies have been conducted on smaller samples or even single galaxies with targeted investigations of outflow dynamics and AGN feedback \citep{fischer2016gemini, shimizu2019multiphase, meena2023investigating}. Such works have laid a foundation for understanding AGN feedback mechanisms and the groundwork for this analysis to expand upon.  

In this manuscript, we present a catalog of Seyfert type AGN aggregated from archival SINFONI and OSIRIS observations. We present both the efforts to aggregate and generate data products of value to the research community as well as the utilization of the dataset to investigate velocity dispersion ($\sigma$) and surface brightness of the gaseous material circumnuclear to the AGN. The Local Universe Near Infrared Seyfert (LUNIS)-AGN Catalog and associated data products highlighted in this work have been made publicly available to stimulate continued AGN research. The primary goals of this work are to:

\begin{enumerate}
\item Create the largest catalog of NIR IFU spectroscopic data for Seyfert type AGN ever aggregated from archival high spatial resolution 3-D datasets.
\item Enhance future AGN studies with easy public access to high quality NIR datacubes and associated data products.
\item Take a first look at exploiting the increased statistics afforded by this larger dataset to uncover fundamental trends in $\sigma$, line luminosity, and surface brightness with fundamental AGN properties. 
\end{enumerate}

Through our analysis of the catalog sample, we aim to better understand the role the AGN play in the excitation of circumnuclear gas and whether or not the circumnuclear gas is significantly influenced by AGN feedback. To do so, we investigate the surface brightness and $\sigma$ of the gas circumnuclear to the AGN and how these properties scale with radial distance, Seyfert classification, and ultrahard X-ray luminosity. We investigate ionized gas emission associated with outflows and AGN activity ([Si VI] and Br-$\gamma$) as well as molecular gas emission (H$_2$ 1-0 S(1)). Our analysis not only attempts to analyze the extent, prevalence and behavior of these gases around the AGN, but also assess the importance of AGN radiative feedback as an excitation mechanism and the interplay between ionized outflows and the molecular gas circumnuclear to the AGN. We aim to accomplish this through comparison of the surface brightness and $\sigma$ both within the central 200 pc nuclear region as well as a how these parameters scale with radial distance from the nucleus for each of the measured emission lines. 

The organization of this paper is as follows: Section 2 presents the catalog, sample selection process, data reduction and data products generated; section 3 presents sample characteristics, subsampling methodology, and techniques for sample analysis of the AGN circumnuclear region across the catalog; section 4 presents our data and observed trends in surface brightness and $\sigma$ across the catalog, as well as our interpretations; and in section 6 we provide a summary of data products and analysis.  

\section{Sample Selection and Data} \label{sec:Sample Selection and Obs}

We present a catalog of NIR IFU data of 88 Seyfert AGN. Relevant IFU observations were obtained with the OSIRIS (OH-Suppressing InfraRed Imaging Spectrograph; \citealt{larkin2006osiris}) instrument of the Keck Observatory and SINFONI (Spectrograph for INtegral Field Observations in the Near Infrared; \citealt{eisenhauer2003sinfoni}) instrument of the ESO’s Very Large Telescope (VLT). We have included archival data from SINFONI/VLT observations available in the ESO Science Archive Facility as of December 2021. This sample is then complemented by suitable OSIRIS/Keck observations available as part of the Keck OSIRIS Nearby AGN survey (KONA; \citealt{muller2018keck}). For object selection, we screened objects identified as AGN in SIMBAD \citep{wenger2000simbad} and select all \textit{K}- (OSIRIS: 1.965-2.381 $\mu$m; SINFONI:  1.95-2.45 $\mu$m) or \textit{H}+\textit{K} (SINFONI: 1.45-2.45 $\mu$m) -band data with an imposed redshift cutoff (z $<$ 0.020; Distance $<$ 83.5 Mpc; a maximum spatial scale of  400 $\frac{pc}{''}$). We accepted data of any plate scale which had a minimum exposure time of 10-minutes as to ensure an acceptable signal-to-noise (S/N) ratio of the emission lines of interest in a “typical” Seyfert galaxy based on previously published line fluxes. With our imposed redshift cutoff, the median full width half max (FWHM) of the point spread function (PSF) across the catalog sample is 39 pc. Our sample was further pruned by comparing object classification to that listed in the NASA/IPAC Extragalactic Database (NED)\footnote{The NASA/IPAC Extragalactic Database (NED) is operated by the Jet Propulsion Laboratory, California Institute of Technology, under contract with the National Aeronautics and Space Administration.} to ensure the galaxy host an AGN and to remove any galaxies known to be hosting a double nucleus.

The sample has a mean redshift of z $\approx$  0.008, median redshift of z $\approx$ 0.007, and a standard deviation of $\approx$ 0.005 (Figure \ref{fig1}). The sample includes a range of the Seyfert types defined by \cite{veron2010catalogue}. In this classification, consistent with AGN unification models, Seyfert type 1s exhibit both narrow and broad Balmer emission lines with relatively high $\sigma$ $>$ 10$^3$ km/s and Seyfert type 2s display only narrow line emission. Seyferts with Balmer lines intermediate between typical Sefyert type 1s and 2s are assigned a fractional classification based on the relative strength of the H-$\beta$ lines compared to [OIII]$\lambda$5007 fluxes with decreasing ratios from Seyfert type 1.2 to 1.9. In addition, a classification of Sefyert type 1i is used for cases where there is a highly reddened broad line region, Seyfert type 1h where broad lines present only in polarized light (also known as hidden broad line region sources), and cases exhibiting low-ionisation nuclear emission line region (LINER) are assigned a Sefyert type 3 classification (see also \citealt{heckman1980optical, malkan2017emission, osterbrock1977spectrophotometry, osterbrock1981seyfert, winkler1992variability, peterson2006broad}).

The largest subset of the sample are Seyfert 2s ($\approx$ 37\%), with Seyferts 1-1.9, 1h, and 1i ($\approx$ 32\%), LINERs ($\approx$ 25\%), and Seyfert 1s ($\approx$ 6\%) comprising the rest. Some galaxies have multiple independent datacubes, from more than one observation by the same instrument or from both instruments. In such instances, we select datacubes for inclusion based on recency of observation and data quality. Both the SINFONI and OSIRIS instruments are equipped with adaptive optics (AO) capabilities which provide enhanced spatial resolution Near-infrared (1-2.5 µm) IFU data \citep{eisenhauer2003sinfoni, larkin2006osiris}. Out of the 88 datasets, 72 (16) were observed with (without) AO (Table \ref{Table_1}). 

All data collected with OSIRIS/Keck was reduced using the OSIRIS Data reduction pipeline. The majority of the SINFONI/VLT data were reduced using SPRED, a custom package developed at the Max Planck Institute for Extratrerrestrial Physics for analysis of SINFONI data \citep{abuter2006sinfoni}.  As indicated in the headers of relevant datacubes, others were reduced using ESO’s SINFONI data reduction pipeline (Version 3.3.3) and run with ESOReflex \citep{freudling2013automated}. In all cases the data reduction performed all the standard reduction steps needed for NIR spectra as well as additional routines necessary to reconstruct the datacubes. We further applied routines designed to improve the subtraction of OH sky emission lines following the approach outlined in \cite{davies2007method}. Telluric correction and flux calibration were carried out typically using A- and B-type stars.  The typical calibration uncertainty in flux as determined from the standard deviation of aperture photometries of the individual observations with SINFONI/VLT (OSIRIS/Keck) is about 10\% (15-20\%) before combining to create a final datacube. In some SINFONI datacubes (indicated in Table \ref{Table_1}), telluric corrections did not remove the Br-$\gamma$ absorption in the standard star, thus an artificial spectral feature was included at 2.1661 $\mu$m. In these cases a linear interpolation has been applied to the affected region in each affected datacube and associated integrated spectra.

Object redshifts were calculated using the centroid of the H$_2$ 1-0 S(1) emission line peak in the nuclear integrated spectrum. For additional quality control, calculated redshifts for each object were compared to previously published values \citep{burtscher2015obscuration, muller2018keck} as well as the NED$^1$ database. Object distances were taken from previous literature when available; \cite{burtscher2015obscuration} and \cite{muller2018keck} provided distance measurement information for roughly 80$\%$ of the datasets within the catalog. Where necessary, the luminosity distance was computed using the object redshift via \cite{wright2006cosmology} using the parameters $\Omega_{matter}$ = 0.27, $\Omega_{vacuum}$ = 0.73, and H$_0$ = 73 $\frac{km}{s*Mpc}$. Distance scales (pc/pixel) were calculated geometrically using the object distance and the plate scale of each observation.  The Seyfert type categorization for each object was determined based on previous literature and cross referenced with NED. Details for the full AGN sample are provided in Table \ref{Table_1}.

\begin{deluxetable*}{cccccccccc}
\tabletypesize{\scriptsize}
\tablewidth{0pt} 
\tablecaption{Key information pertaining to the catalog sample.  \label{Table_1}}
\tablehead{
\colhead{Galaxy} & \colhead{Instrument}& \colhead{Seyfert Type$^m$} & \colhead{Redshift} &
\colhead{Plate Scale} & \colhead{Distance} & \colhead{pc/$\arcsec$}& \colhead{$L_{14-195keV}$$^g$} & \colhead{N$_H$} & \colhead{PSF$^k$} \\
\colhead{} & \colhead{} &
\colhead{} & \colhead{} & \colhead{$\arcsec$/pix} &\colhead{Mac} &\colhead{} & \colhead{log erg s$^{-1}$} & \colhead{log cm$^{-2}$} & \colhead{pc}
} 
\colnumbers
\startdata 
Circinus$^{l}$		&	SINFONI	&	1h$^b$	&	0.0014	&	0.0125	&	4.2$^b$	&	20	&	41.76	&	        &	6	\\	
ESO	137-G034	    &	SINFONI	&	2	    &	0.00938	&	0.05	&	38.2	&	185	&	42.62	&	24.36	&	34	\\
ESO	428-G14	        &	SINFONI	&	2$^b$	&	0.0056	&	0.05	&	24.3$^b$ &	118	&	42.22$^b$	&	   &	34	\\	
ESO	548-81$^{l}$	&	SINFONI	&	1$^b$	&	0.0145	&	0.05	&	58.5$^b$	&	284	&	43.23	&	20	&	62	\\
IC	1459$^{l}$	&	SINFONI	&	3$^{a,b}$	&	0.006	&	0.05	&	28.4$^b$	&	138	& 42.29$^b$	&	&	52	\\
IC	2560	&	SINFONI	&	2$^a$	&	0.00984	&	0.05	&	40.7$^f$	&	197	&	42.36$^h$	&	&	39	\\	
IC	4296	&	SINFONI	&	3$^a$	&	0.0133	&	0.125	&	55.2$^f$	&	268	&	&	&	182	\\		
IC	4329a	&	OSIRIS	&	1.2$^a$	&	0.0157	&	0.035	&	66.2$^d$	&	321	&	44.14	&	21.52	&	39	\\
IC	5063*$^{l}$	&	SINFONI	&	1h$^{a,b}$	&	0.0113	&	0.05	&	41.7$^b$	&	202	&	43.15	&	23.56	&	58	\\
IRAS	01475-0740	&	OSIRIS	&	2$^d$	&	0.0168	&	0.05	&	73$^d$	&	354	&	42.42$^d$	&	&	34	\\	
M	87$^{l}$	&	SINFONI	&	3$^{a,b}$	&	0.0042	&	0.05	&	17.1$^b$	&	83	&	&	&	28	\\		
MCG	-44	&	SINFONI	&	1i$^b$	&	0.00869	&	0.05	&	39.5$^b$	&	168	&	43.59	&	22.18	&	43	\\
MCG	-06-30-015*	&	SINFONI	&	1.2	&	0.00808	&	0.125	&	26.8	&	130	&	42.71	&	20.85	&	81	\\
MRK	1066	&	KONA	&	2$^{a,c}$	&	0.0122	&	0.05	&	49.5$^d$	&	240	&	42.47	&	&	23	\\	
MRK	1210	&	OSIRIS	&	2$^d$	&	0.0136	&	0.035	&	58$^d$	&	281	&	43.36	&	23.4	&	42	\\
MRK	573	&	OSIRIS	&	2a	&	0.01635	&	0.035	&	71.4$^d$	&	346	&	43.34$^d$	&	&	45	\\	
MRK	766	&	OSIRIS	&	1$^{a,c}$	&	0.0122	&	0.05	&	54.9$^d$	&	266	&	42.98	&	20.32	&	25	\\
NGC	1052$^{l}$	&	SINFONI	&	3h$^{a,b,c}$	&	0.005	&	0.05	&	20$^b$	&	97	&	42.18	&	22.95	&	25	\\
NGC	1068$^{l}$	&	SINFONI	&	1h$^{b,c}$	&	0.0038	&	0.05	&	14.4$^b$	&	70	&	41.97	&	25	&	13	\\
NGC	1097$^{l}$	&	SINFONI	&	3b$^{a,b}$	&	0.0042	&	0.05	&	17$^b$	&	82	&	&	&	19	\\		
NGC	1194	&	OSIRIS	&	1.9$^a$	&	0.0128	&	0.05	&	55.3$^d$	&	268	&	43.12	&	24.18	&	28	\\
NGC	1320	&	OSIRIS	&	2$^{a,d}$	&	0.00948	&	0.05	&	36.7$^d$	&	178	&	42.33	&	&	21	\\	
NGC	1365	&	SINFONI	&	1.8$^{a,b}$	&	0.00577	&	0.05	&	17.9$^b$	&	87	&	42.39	&	22.21	&	24	\\
NGC	1386$^{l}$	&	SINFONI	&	1i$^{a,b}$	&	0.0029	&	0.05	&	16.2$^b$	&	79	&	41.8$^b$	&	&	31	\\	
NGC	1566$^{l}$	&	SINFONI	&	1.5$^{a,b}$	&	0.005	&	0.05	&	12.2$^b$	&	59	&	41.54	&	20	&	26	\\
NGC	1614*$^{l}$	&	SINFONI	&	2$^a$	&	0.0156	&	0.125	&	64.9$^f$	&	315	&	43.05$^h$	&	&	287	\\
NGC	1667	&	OSIRIS	&	2$^{a,d}$	&	0.0153	&	0.035	&	65$^d$	&	315	&	40.46$^d$	&	&	110	\\	
NGC	1672*	&	SINFONI	&	2$^a$	&	0.0049	&	0.125	&	20.2$^f$	&	98	&	41.68$^h$	&	&	152	\\
NGC	1808*	&	SINFONI	&	2$^a$	&	0.0036	&	0.125	&	14.8$^f$	&	72	&	41.76$^b$	&	&	64	\\
NGC	2110	&	SINFONI	&	1i$^{a,b,c}$	&	0.00802	&	0.05	&	35.6$^b$	&	173	&	43.7	&	22.94	&	52	\\
NGC	253*	&	SINFONI	&	2$^a$	&	0.00077	&	0.125	&	3.2$^f$	&	16	&	40.54$^h$	&	&	9	\\	
NGC	262	&	OSIRIS	&	2$^d$	&	0.0142	&	0.035	&	64.1$^d$	&	311	&	43.85	&	23.12	&	78	\\
NGC	2911$^{l}$	&	SINFONI	&	3$^{a,b}$	&	0.0102	&	0.05	&	52$^b$	&	252	&	&	        &	174	\\		
NGC	2974$^{l}$	&	SINFONI	&	2$^{a,b}$	&	0.0066	&	0.05	&	24.7$^b$	&	120	&	42.02$^b$	&	&	41	\\	
NGC	2992	&	SINFONI	&	1i$^{a,b,c}$	&	0.00797	&	0.05	&	31.6$^b$	&	153	&	42.59	&	21.72	&	62	\\
NGC	3081	&	SINFONI	&	1h$^{a,b,c}$	&	0.008	&	0.05	&	26.5$^b$	&	128	&	42.84	&	23.91	&	46	\\
\enddata
\tablecomments{
* values indicate datasets collected without AO. a: Referenced from the NED database; b: \cite{burtscher2015obscuration}, X-ray luminosity values sourced from this manuscript were adjusted for distance; c: \cite{riffel2017gemini}; d: \cite{muller2018keck};  e: \cite{dumont2019surprisingly}; f: \cite{wright2006cosmology}; g: 157 Month \textit{Swift-BAT} unless otherwise noted; h: The AGN NIR flux was estimated using the CO equivalent width radial profile as described in \cite{burtscher2015obscuration} and the log(L$_{14-195keV}$) was estimated using the linear relation presented by \cite{burtscher2015obscuration}; i: the AGN MIR flux from  \cite{asmus2014subarcsecond} and the log(L$_{14-195keV}$) was estimated using the linear relation presented by \cite{burtscher2015obscuration}.; j: Column Density measurements taken from \cite{ricci2017bat}; k: Represents the projected on sky FWHM of the PSF.; l: Datasets in which the artificial emission line at 2.2661 $\mu$m is present in associated datacubes.; m: Type 3 classification is equivalent to LINER.}
\end{deluxetable*}

\subsection{General Characteristics of the LUNIS-AGN Catalog Sample}\label{Catalog_Characteristics}
\begin{figure*}[ht]
\plotone{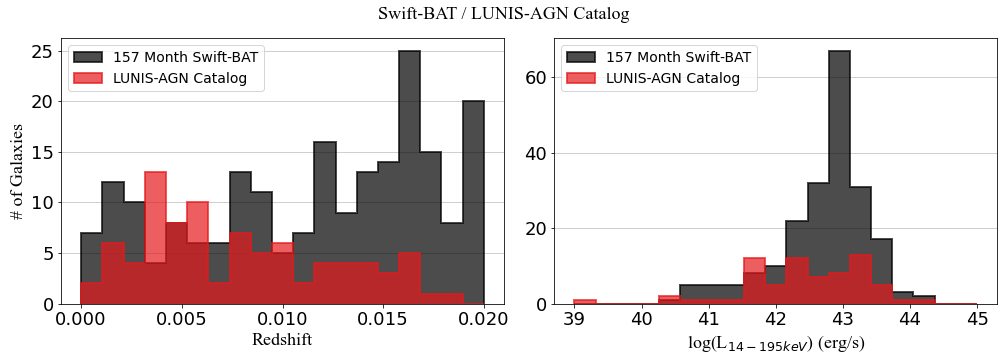}
\caption{Redshift histogram (left) and X-ray luminosity (right) of the AGN catalog overlain on the \textit{Swift-BAT} sample truncated at z = 0.02.   
\label{fig1}}
\end{figure*}

We compare the sample to the unbiased \textit{Swift-BAT} sample to assess any potential biases and for comparison of general sample characteristics. Figure \ref{fig1} compares the AGN catalog to the 157-month \textit{Swift-BAT} database in terms of redshift as well as X-ray luminosity. 53\% of the catalog sample is included in the 157-month \textit{Swift-BAT} database (Lien et al. 2024, in prep; \citealt{baumgartner201370, ricci2017bat}).  L$_{14-195kev}$ X-ray luminosites were calculated from the observed flux reported in the 157 Month \textit{Swift-BAT} survey where available. Where unavailable, the L$_{14-195keV}$ luminosity was estimated from the AGN NIR Luminosity (L$_{NIR}^{AGN}$) via the  relationship described by \cite{burtscher2015obscuration}. Estimates for AGN X-ray luminosity were obtained for all non-LINER AGN and are reported in Table \ref{Table_1}. 

Comparing to the 157 month \textit{Swift-BAT} sample  (mean redshift z $\approx$ 0.012 $\pm$ 0.0004), the redshift distribution of AGN within the LUNIS-AGN catalog has a tendency towards lower redshift (mean redshift z $\approx$ 0.008 $\pm$ 0.0005) (Figure \ref{fig1}).  The distribution of X-ray luminosity within the catalog is similar to that of the BAT survey, with a mean log(L$_{14-195keV}$) X-ray Luminosity of 42.45 $\pm$ 0.094 erg s$^{-1}$ compared to 42.69 $\pm$ 0.052 erg s$^{-1}$ for the BAT sample.  Utilizing a Kolmogorov-Smirnov (K-S) test to compare the sample redshift and X-ray luminosity distribution of the LUNIS-AGN catalog and the 157 Month \textit{Swift-BAT} sample, we find that sample distributions do differ in a statistically significant way. This difference in sample distributions is likely driven by the subsample of AGN within the LUNIS-AGN catalog which are not represented in the 157 month \textit{Swift-BAT} survey. These AGN are likely not represented in \textit{Swift-BAT} due to their lower X-ray luminosity. The mean redshift and estimated log[L$_{14-195keV}$] X-ray luminosity of this sub-sample is $\approx$ 0.007 $\pm$ 0.0006 and $\approx$ 41.74 $\pm$ 0.13 erg s$^{-1}$ respectively.  We conclude that the catalog distribution does not represent the local AGN population as sampled by the \textit{Swift-BAT} survey, but instead spans a more extensive population including lower luminosity AGN.

\subsection{Data Products} \label{subsec:Data_Products}

All datacubes and associated data products within this catalog are publicly available at \url{https://doi.org/10.7910/DVN/HHYIJ2}.  If making use of the provided data products please cite this publication. Data products provided as part of this catalog include:
\begin{itemize}
    \item 3D \textit{K}-band or \textit{H+K} -band (where available)  NIR IFU datacubes for each galaxy.
    \item Associated integrated spectrum for the central 200 pc region around the galactic nucleus. 
    \item 2D maps of the flux distribution, velocity field, and $\sigma$ for H$_2$ 1-0 S(1), [Si VI], and Br-$\gamma$ emission lines.
    \item Azimuthally averaged luminosity (L$_{\lambda}$ $\cdot$ pc$^{-2}$) profile for the H2 1-0 S(1), [Si VI], and Br-$\gamma$ emission lines 
    \item Orientation of [SI VI] emission as a function of radius as well as an estimate of extent of [SI VI] emission; the position angle and extent of the emission at the FWHM and 10\% max [SI VI] flux is provided.

\end{itemize}

\begin{figure*}[ht]
\plotone{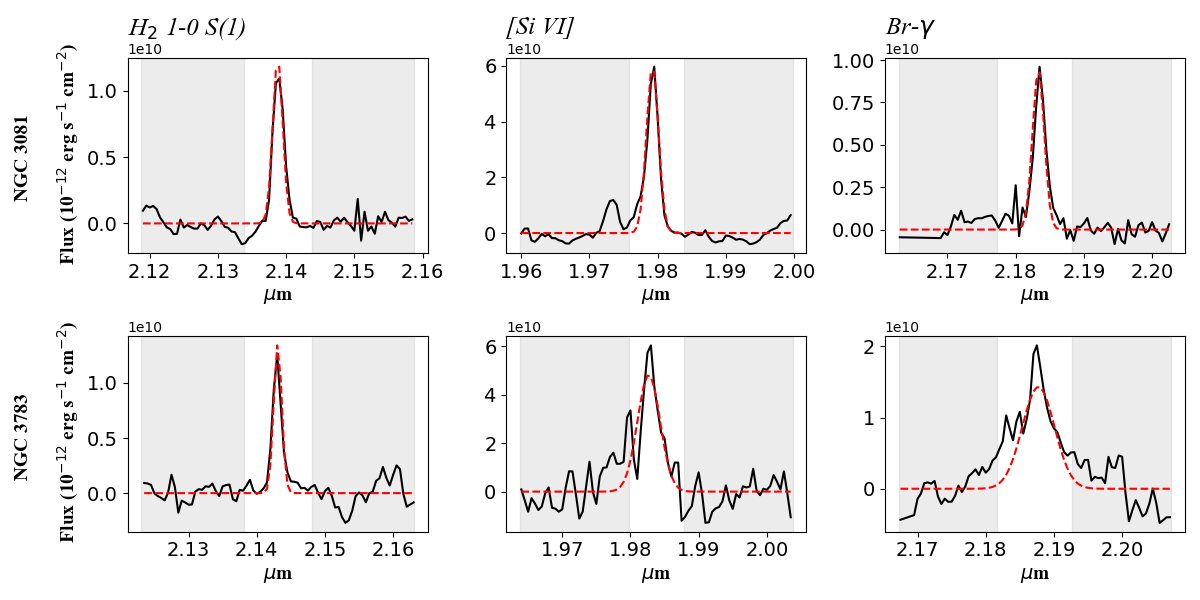}
\caption{Example Gaussian fit results for the central 200 pc integrated spectrum for NGC 3081 (Seyfert 1h; top panel) and NGC 3783 (Seyfert 1.5; Bottom panel). The shaded regions indicate spectral range used for the linear continuum subtraction. 
\label{Gaussian Fits Example}}
\end{figure*}

The 2D maps of the emission line flux, velocity, and $\sigma$ were generated using a custom IDL code (Linefit; \citealt{davies2007method}) in which a single component Gaussian profile was fit to the emission line of interest and linear fit to characterize the continuum.  Table \ref{Table 2} provides the spectral range used in the Gaussian fit to the emission line at each spaxel as well as the spectral windows on either side of this used to establish the continuum. These spectral ranges were chosen pragmatically to provide adequate range to fit the continuum, but not enough to risk inclusion of additional emission lines. While our continuum windows were selected to mitigate the inclusion of other lines, the Br-$\delta$ emission line at a 1.944 $\mu$m is precariously close to the [Si VI] emission line and there is potential for contamination. Therefore there is potential for improving maps by optimizing the continuum and fit windows for each target or by using multi-component fitting algorithms. Additionally, no attempt has been made to separate potential narrow and broad components in the Br-$\gamma$ emission. Seyfert 1 galaxies, where a broad Br-$\gamma$ component is expected, will therefore have some level of contamination in the values reported within a central region depending on the spatial resolution of the data (typically $\approx$ 50 pc; Table \ref{Table_1}). Figure \ref{Gaussian Fits Example} presents example fits for the 200 pc integrated spectra for two representative objects within the sample, the Seyfert 1.5 galaxy NGC 3081 and the Seyfert 1h galaxy NGC 3783. The presence of the broad component, and impact to the emission line fit for Br-$\gamma$ is apparent for NGC 3783.    

\begin{deluxetable*}{ccccc}
\tabletypesize{\scriptsize}
\tablewidth{0pt} 
\tablecaption{Spectral range parameters utilized for single-Gaussian fitting to the emission lines of interest.  \label{Table 2}}
\tablehead{
\colhead{Emission Line} & \colhead{Central Wavelength ($\mu$m)}& \colhead{Gaussian Fit* ($\mu$m)} &
\multicolumn{2}{c}{Continuum* ($\mu$m)}\\
\cline{1-5}
\colhead{} & \colhead{} &
\colhead{} & \colhead{Left} & \colhead{Right} 
} 
\colnumbers
\startdata 
\textbf{H$_2$ 1-0 S(1)}&  2.1218&-0.004/+0.004 & -0.02/-0.005 & +0.005/+0.02 \\
\textbf{[Si VI] }&  1.9641&-0.004/+0.005 & -0.02/-0.004 & +0.005/+0.02 \\
\textbf{Br-$\gamma$}&  2.1655&-0.0045/+0.006 & -0.02/-0.0055 & +0.0055/+0.02 \\ 
\enddata
\tablecomments{*Gaussian fit and continuum window denote shifts for each respective emission line.}
\end{deluxetable*}

\begin{figure*}[ht]
\plotone{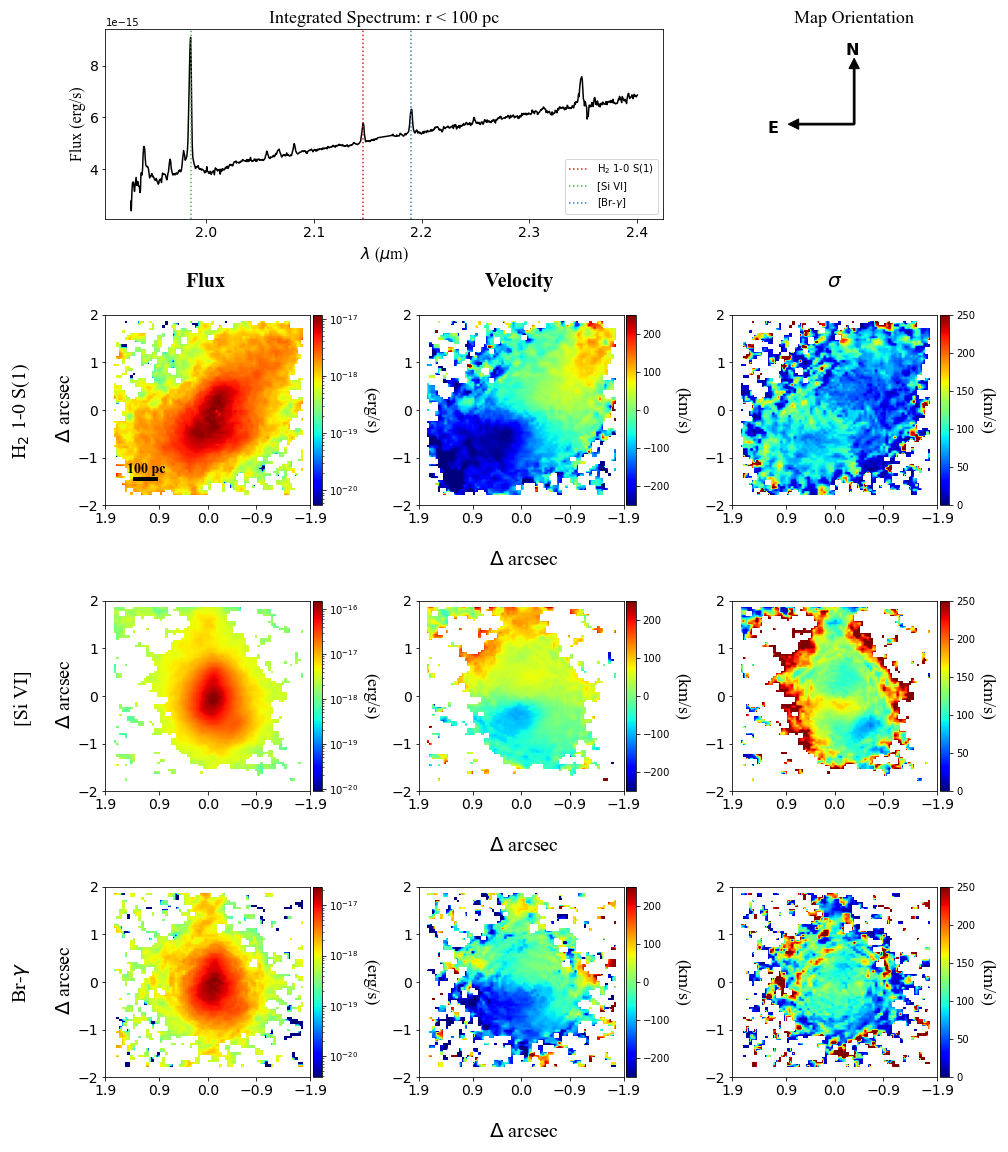}
\caption{Generated data products, in this case for NGCC 3281.  The top panel depicts the integrated spectrum of the central 200 pc. Below are the associated line maps, these include: flux (left), velocity (center), and velocity dispersion (right) maps generated for each H$_2$  1-0 S(1) (2.12 micron), [Si VI] (1.96 microns), and Br-$\gamma$ (2.16 microns) from top to bottom. 
\label{fig2}}
\end{figure*}

The integrated spectrum and 2D flux, velocity, and $\sigma$ maps have been generated for each galaxy in the sample. An example set of these data products is shown in Figure \ref{fig2}. Emission line maps presented in this figure have a basic data masking procedure which consists of the removal of pixels with extremely high velocity dispersion ($\sigma$ $>$ 800 km/s), which we find is indicative of a poor fit to the emission line profile, and application of a S/N flux threshold in which pixels with S/N $<$ 3 were removed. We provide, as part of our suite of data products, datasets without masking applied and a S/N mask which can be applied to the dataset.  

\begin{figure*}[ht]
\plotone{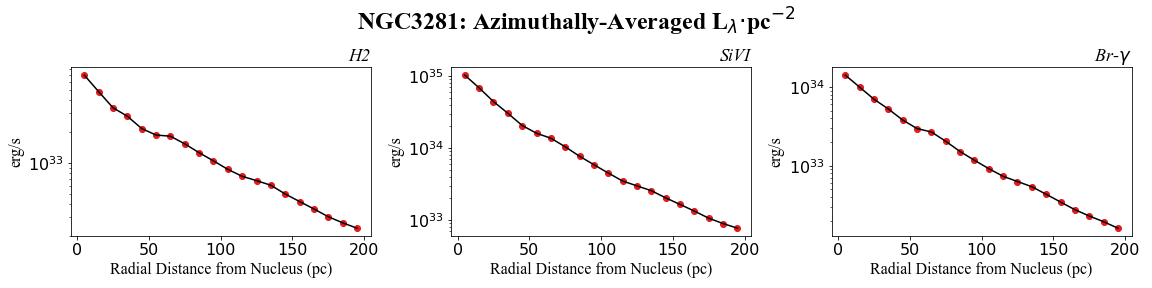}
\caption{Example of luminosity azimuthal average plots generated for each of the galaxies in the catalog. Displayed in this figure is the luminosity azimuthal average generated for the H$_2$  1-0 S(1), [Si VI], and Br-$\gamma$ emission lines for NGC 3281. Red points represent the midpoint of each 10 pc circular annuli over which the luminosity was averaged; representing the average pixel luminosity at a given radii.   
\label{fig3}}
\end{figure*}

Additionally, we have generated and provide plots of the azimuthally-averaged L$_{\lambda}$$\cdot$pc$^{-2}$ out to 200 pc from the galactic nucleus. Figure \ref{fig3} depicts an example set of plots of the luminosity azimuthal average for each emission line investigated. A similar set of plots has been generated for each galaxy in the catalog and has been made publicly available. 

\begin{figure*}[ht]
\plotone{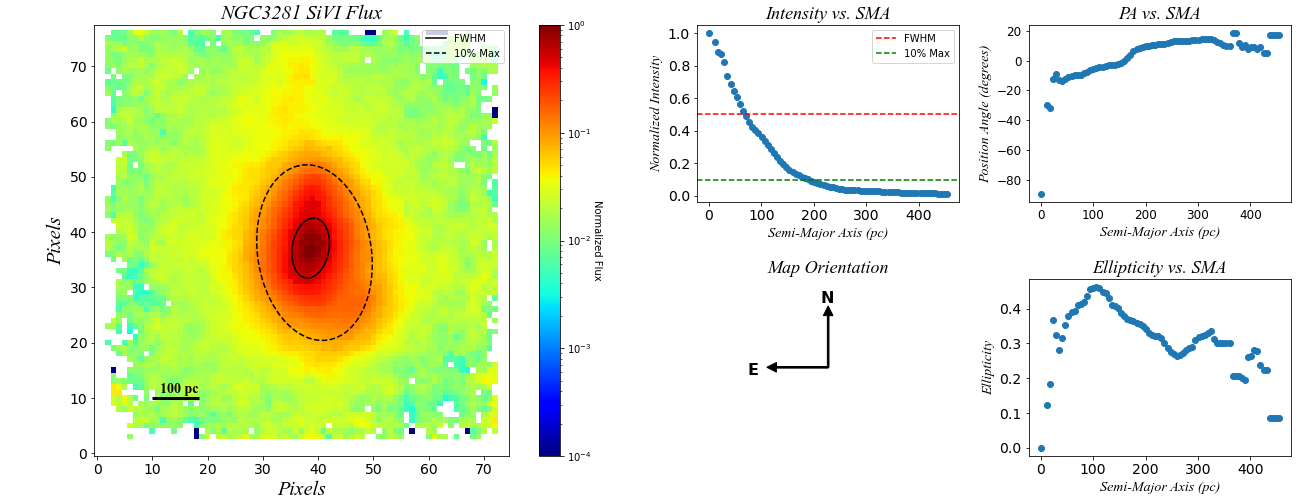}
\caption{Left panel depicts the [Si VI] normalized flux map for NGC 3281 with elliptical fits for the 10\% max flux (dashed line) and the FWHM (solid line). Sub-panels to the right depict the map orientation (bottom left), normalized intensity (top left),  position angle (top right) and Ellipticity (bottom right) of the ellipse as a function of semi major axis length. The horizontal lines indicate the values of the FWHM (red dashed line) and the point of 10\% max flux (green dashed line).  
\label{fig4}}
\end{figure*}

Given the association of [Si VI] emission with AGN outflows, this emission line was analyzed to investigate extent and orientation of the emission. This detailed analysis was applied only to those galaxies in which the [Si VI] emission was found to be above a S/N threshold of 3 in the central 200 pc integrated spectrum (78 of the 88 galaxies within the catalog). Using the  AstroPy-affiliated package PHOTUTILS, elliptical isophotes were fit to the observed [Si VI] emission to characterize its morphology. An elliptical fit was achieved for 58 of the 88 AGN within the catalog. Where a meaningful fit could be achieved, the position angle, ellipticity, and normalized intensity was examined as a function of semi-major axis length (see Figure \ref{fig4} for an example). 

\begin{figure*}[ht]
\plotone{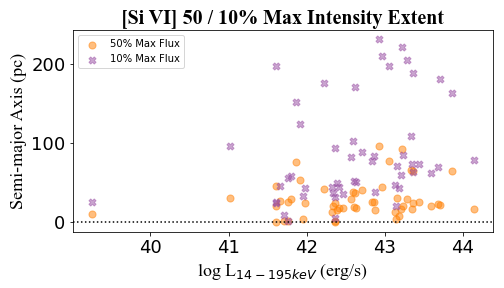}
\caption{Points represent the semi major axis length of the elliptical isophote fit to the [Si VI] emission. A value of zero indicates that the flux value decreased below 50\%/90\% within a single pixel extent.    
\label{fig5}}
\end{figure*}

Relevant information regarding the position angle (PA), ellipticity and semi-major axis length (SMA) of the elliptical fits achieved at the 50$\%$ and 10\% max [Si VI] emission intensity is provided in Appendix A. All data generated as part of this analysis pertaining to elliptical fitting, including extent and position angle of fit, have been made publicly available as part of the Seyfert AGN catalog. The primary limiting factor in achieving meaningful elliptical fits was data quality.  No dependence of the 10\% max flux or FWHM extent of [Si VI] emission with X-ray luminosity or Seyfert type was observed in this analysis (Figure \ref{fig5}).  

\section{Characteristics of the Circumnuclear Gas and its Relationship to the AGN} \label{sec:Section3}

Here we highlight the motivation and methodology for parsing the LUNIS-AGN catalog into data subgroups and outline the methodology we employ for analysis of the circumnuclear region. We have taken the first steps at utilizing the increased statistics supported by this relatively large archival sample (88 AGN included within the catalog) to investigate potential trends in the $\sigma$ and surface brightness of key gas tracers (H$_2$ 1-0 S(1), [Si VI], and Br-$\gamma$) with fundamental AGN properties (Seyfert type, obscuring column density, and X-ray luminosity). These emission lines were selected for analysis because they are strong in typical Seyfert galaxies and trace the warm molecular interstellar medium (ISM; H$_2$  1-0 S(1)) as well as tracing ionized gas often associated with AGN outflows (Br $\gamma$ and [Si VI]; e.g. \cite{riffel2020ionized}.

\subsection{Defining Sample Subgroup: AGN type and X-Ray Luminosity} \label{subsec:tables}

\begin{figure*}[ht]
\plotone{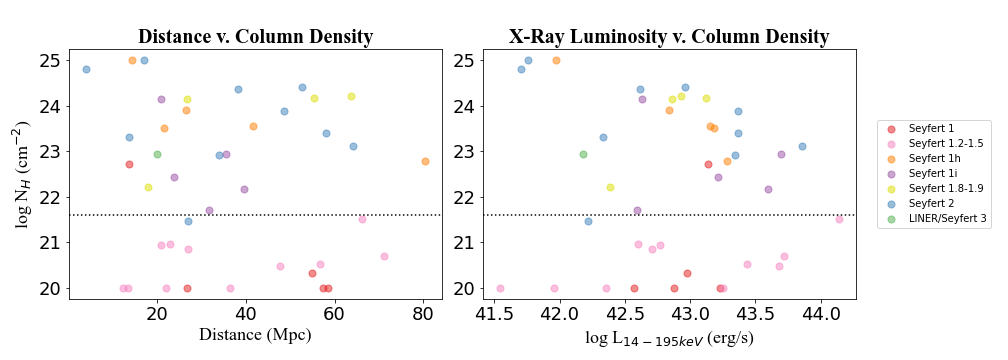}
\caption{Measured column density vs. distance.  Note, a column density value of 20 represents an artificial value reported due to the flux being below the sensitivity limits of the survey. The dotted line indicates the general separation between Seyfert subgroups.  
\label{fig6}}
\end{figure*}

To better split the catalog datasets and develop a pragmatic binning for data analysis, we examine Seyfert subgroups as a function of X-Ray luminosity (L$_{14-195 keV}$) and estimated column density (N$_H$). Obscuration is thought to be a key factor in driving observed differences in the observed Balmer lines and therefore integrally tied to Seyfert classification (see Section \ref{sec:Sample Selection and Obs}). In an effort to better quantify obscuration of the AGN, we utilize measurements of the obscuring column density as reported by \cite{ricci2017bat}. Column density measurements are available for 41 of the 88 AGN within the catalog. Figure \ref{fig6} depicts log[N$_{H}$] vs. distance and L$_{14-195keV}$ luminosity for the subset of the catalog for which N$_H$ measurements are available, partitioned by Seyfert type.  While we observe no correlation with N$_H$ and measured L$_{14-195 keV}$ within the sample, we note that the datasets exhibit a clear separation between Seyfert types 1, 1.2, 1.5, and other types at log(N$_{H}$/cm$^{-2}) \approx$ 21.7. 

\subsubsection{Seyfert Type Binning - Obscuration} \label{Seyfert_Binning}

\begin{figure*}[ht]
\plotone{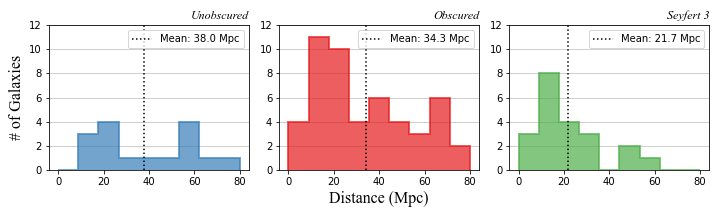}
\caption{Distance histogram of applied Seyfert classifications  
\label{fig7}}
\end{figure*}

We find that N$_H$ is well correlated with Seyfert type (Figure \ref{fig6}), as has been found in previous studies and is consistent with the AGN unification model \citep{hickox2018, malkan2017emission}. We take this as motivation to combine, for the purposes of this analysis, the Seyfert 1 and Seyfert 1.2-1.5 galaxies as a single subgroup, called here “unobscured Seyferts” and Seyfert 1.8-1.9, 1h, 1i, and Seyfert 2 into a single subgroup, called here “obscured Seyferts". Utilization of these subgroups provides the benefit of tethering our analysis of Seyfert AGN to the fundamental physical property of obscuration. For the sample analysis highlighted in this manuscript, AGN will be categorized into these applied Seyfert/obscuration groups.  The distance distribution for each of these Seyfert subgroups is shown in Figure \ref{fig7}. The mean distances are 38.0 ± 5.0 pc, 34.3 ± 3.0 pc, 21.7 ± 3.0 pc for the unobscured Seyfert, obscured Seyfert, and Seyfert3/LINER groups, respectively.  While the mean distance of the unobscured and obscured subgroups is similar, the LINER AGN tend to be significantly closer and will therefore generally these datasets will be of a higher spatial scale relative to the those of unobscured and obscured AGN.

\subsubsection{X-Ray Luminosity Binning - AGN Energetics}\label{Xray_Binning}

\begin{figure*}[ht]
\plotone{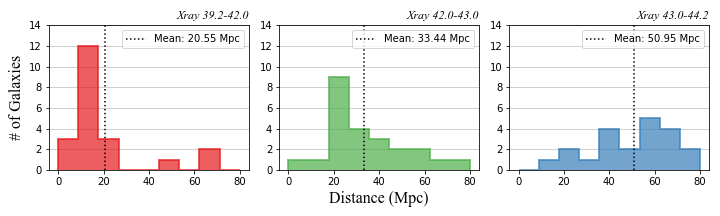}
\caption{Distance histograms of each X-ray luminosity bins utilized in this analysis.   
\label{fig8}}
\end{figure*}

X-ray luminosity binning was also employed to organize data into more statistically significant subgroups for analysis of the circumnuclear region as a function of L$_{14-195 keV}$.  We utilize the observed L$_{14-195 keV}$ as an indicator of the AGN power which theoretically scales with the rate of accretion \citep{trump2011accretion, ricci2017bat}.  Figure \ref{fig6} shows the distribution of N$_H$ as a function of L$_{14-195 keV}$. Measured L$_{14-195 keV}$ range from 10$^{39}$ to 10$^{44}$ erg/s, so for simplicity, data were binned into three intervals chosen to represent low, medium, and high power Seyfert AGN: log (L$_{14-195 keV}$) = 39.2-42.0, 42.0-43.0, and 43.0-44.2 erg/s. The proportion of unobscured AGN within each bin is $\approx$ 10$\%$, 29$\%$, and 32$\%$ respectively. Figure \ref{fig8} depicts the distance histograms for each X-ray luminosity bin. The mean distance for these X-ray bins is 20.5 ± 3.7 pc, 33.4 ± 3.4 pc, and 50.9 ± 3.8 pc for log (L$_{14-195 keV}$) = 39.2-42.0, 42.0-43.0, and 43.0-44.2 erg s$^{-1}$, respectively. We observe a clear trend in which object distance positively correlates with L$_{14-195 keV}$, likely an artifact of the ease of detecting more luminous objects at greater distances relative to AGN of lower luminosity. Therefore, generally, we anticipate higher power AGN within the LUNIS-AGN catalog to have a bias towards decreased spatial scale.

\subsection{Methodology for Characterizing the Circumnuclear Gas} \label{Methods}

Both surface brightness and the average value of $\sigma$ were investigated for various apertures within the circumnuclear regions of the sample for all three emission lines. No correction for the galaxy orientation (e.g. disk inclination) with respect to the line of sight has been applied and thus all distances represent projected distances from the central AGN. We present the average values within r=100 pc of the AGN (central 200 pc; Tables \ref{table 3} and \ref{Table 4}) as well as the azimuthal average using 10 pc circular annuli out to r=200 pc (central 400 pc). Through this analysis, we compare the behavior of measured parameters for each galaxy within the catalog and more generally, assess their dependence on Seyfert classification and ultrahard X-ray luminosity. We analyze the azimuthal average values of $\sigma$ and surface brightness for the purpose of better understanding how these parameters, on average, change as a function of radius as well as Seyfert type (obscuration) and X-Ray luminosity.  We examine our results in terms of mean values for each galaxy, as well as the median of these means across each subset of Seyfert AGN.  

Characterization of the circumnuclear gas was done through analysis of the 2D flux and $\sigma$ maps of H$_2$ 1-0 S(1), [Si VI], and Br-$\gamma$ with all spaxels containing a line flux with a S/N $>$ 3.  Additionally, measurements where the $\sigma$ exceeded 800 $\frac{km}{s}$ were also masked as it was determined that above this threshold a single Gaussian does not represent the line profile due to either an underlying broad component or, in rare cases, a double peaked emission line. As already noted, no attempt has been made to separate potential narrow and broad components in the Br-$\gamma$ emission.  Seyfert 1s, where a broad Br-$\gamma$ component is expected, will therefore have some level of contamination in the values reported within a central region depending on the spatial resolution of the data (typically $\approx$ 50 pc). To alleviate the impact of any data bias pertaining to object distance, i.e. dataset resolution (Section \ref{Xray_Binning}), we utilize apertures of fixed projected sizes in physical units. 

For our azimuthal analysis, our sampling increment of 10 pc annuli is consistent with the mean spatial resolution of the sample ($\approx$ 9.0 pc/pixel). We acknowledge that while our choice of sampling increment is consistent with the mean spatial sampling of the catalog data, the spatial resolution characterized by the FWHM of the PSF is larger with a mean value of approximately 65 pc (Table \ref{Table_1}). As part of our analysis, we have tested utilizing a larger spatial binning consistent with this average spatial resolution and note no difference in analysis outcomes. As such, we maintain use of the 10 pc sampling increment to ensure full sampling of our highest resolution datasets. 

The center of each galaxy was determined by fitting a 2-D Gaussian to the distribution of the total flux of the datacube. As a quality control step, we implement a cutoff value for live pixels within each annuli. Annuli with $<$20\% of pixels live after data masking were discarded to avoid biasing sample measurements towards annuli with overall low S/N.  Subpixel fractional contributions were utilized for spaxels which do not fully fall within the circular apertures. 

\section{Results} \label{sec:Results}

Throughout our analysis, we focus on Unobscured and Obscured Seyfert type AGN. These AGN are of particular interest in the context of AGN unification theories, in understanding AGN inflow and outflow processes, and characterizing the outer extension of the circumnuclear obscuring structure or torus. For completeness, LINER data are presented as it is available, however no interpretation or discussion regarding the behavior of this AGN subgroup is included as it is beyond the scope of this work. In addition, for consistency, galaxies classified as LINER with available X-ray measurements (NGC1052,  IC1459, and NGC4579) were excluded from our analysis of trends with L$_{14-195 keV}$. We probe the circumnuclear region to asses any links between surface brightness and $\sigma$ and the AGN parameters of obscuration and L$_{14-195keV}$. We examine these parameters within the central region as well as how they change with radial distance from the active nucleus. To reiterate, our goal is to understand the role that AGN play in the excitation and kinematics of the gaseous circumnuclear material, both outflowing material traced by ionized gas and molecular rotating gases as traced by H$_2$ 1-0 S(1) emission.  

To identify any underlying fundamental differences in the circumnuclear gas in relation to obscuration or L$_{14-195 keV}$, we consider both the integrated spectra within the central 200 pc (r$<$100pc) and the 2D emission line maps. To visualize results based on our 2D data, including radial distance from the nucleus, where relevant, we utilize $\sigma$ vs. surface brightness space (Figures \ref{fig9} and \ref{fig10}). This space provides the efficiency to view all relevant parameters, including radial distance from nucleus, Seyfert or X-ray subgroups, $\sigma$, and surface brightness within a single plot. These plots summarize results based on the 2D emission line maps for (1) the r $<$ 100 pc mean values of $\sigma$ plotted against the r $ < $ 100 pc surface brightness for each galaxy and (2) the median values of the mean $\sigma$ and surface brightness for radial bins across the entire sample. In addition, radial plots of the $\sigma$ and surface brightness for each of the subsamples with standard error bars have been included in Appendix A (Figures \ref{figA1} and \ref{figA3}).We focus our analysis of $\sigma$ on the 2D maps rather than the integrated spectra to minimize the contribution from beam smearing, which could increase the dispersion due to unresolved velocity gradients. In plots in the left column of Figures \ref{fig9} and \ref{fig10}, we measure the mean $\sigma$ from spaxels within the central 200pc, which is at least two times the PSF FWHM (and typically more than three times the PSF FWHM for this sample) and thus well outside of significant PSF effects. Future modeling considering the velocity field is necessary to properly assess the level of impact beam smearing could have on measurements within the spatial resolution of the data. Given the statistical approach taken in this study, and the significant variation across Seyfert subgroups, we do not anticipate this effect impacts our conclusions.

\subsection{Emission Line Detections and Flux Distribution} \label{General_Results}

Table \ref{Table_A1} presents the luminosity measured from the central 200 pc integrated spectra as well as the measured half width half max (HWHM) of the flux distribution for each emission lines measured. Across the 88 objects included in the LUNIS-AGN sample, we report a reliable detection within the central 200 pc integrated spectrum for $\approx$ 92$\%$, 73$\%$, and 82$\%$ of the datasets for the H$_2$ 1-0 S(1), [Si VI], and Br-$\gamma$ respectively. To first order, the flux distribution of these emission can be understood via the HWHM of the azimuthally-averaged L$_{\lambda}$$\cdot$pc$^{-2}$ (obtained from the 10 pc annuli as described in section \ref{Methods}).  HWHM values were obtained simply by taking the distance between the annuli of the maximum and the 10 pc annuli at which the average L$_{\lambda}$$\cdot$pc$^{-2}$ reached half of the maximum. A visual check was employed to ensure measurement quality and eliminate cases where the maximum L$_{\lambda}$$\cdot$pc$^{-2}$ is offset from the AGN location. The mean HWHM values across the catalog sample, as well as for each binning regime, is presented in Table \ref{Table HWHM}. We find that, across the full catalog sample as well as for each individual binning regime, the H$_2$ 1-0 S(1) luminosity HWHM to be greater than that of [Si VI] and Br-$\gamma$, which have similar HWHM values. This result suggests that the ionized gas associated with [Si VI] and Br-$\gamma$ is more centrally concentrated than the molecular hydrogen gas. This result is consistent with previous studies which examined the distribution of molecular and ionized gas around the AGN \citep{schonell2019gemini, riffel2021agnifs}. This difference is least pronounced in the lowest L$_{14-195 keV}$ bin, which we attribute to the scaling of [Si VI] and Br-$\gamma$ line luminosity with AGN power (Section $\ref{XrayLum_vs_LineLum}$).

\begin{deluxetable*}{cccc}
\tabletypesize{\scriptsize}
\tablewidth{0pt} 
\tablecaption{Emission Line Flux Distribution: Line Luminosity HWHM.  \label{Table HWHM}}
\tablehead{
\colhead{} & \colhead{H$_2$ 1-0 S(1)} &
\colhead{[Si VI]} & \colhead{Br-$\gamma$} \\
\colhead{} & \colhead{pc} &
\colhead{pc} & \colhead{pc} 
} 
\colnumbers
\startdata 
\textbf{Full Sample}& 24.6 $\pm$ 0.2& 21.2 $\pm$ 0.2& 21.1 $\pm$ 0.2\\ 
\cline{1-4}
\textbf{Unobscured}& 18.1 $\pm$ 0.7 & 13.1 $\pm$ 0.3 & 14.6 $\pm$ 0.4\\ 
\textbf{Obscured}& 26.4 $\pm$ 0.4 & 18.5 $\pm$ 0.2 & 20.6 $\pm$ 0.3\\ 
\cline{1-4}
\textbf{39.2 $<$ log(L$_{14-195keV}$) $\leq$ 42.0}& 20.5 $\pm$ 0.5& 17.8 $\pm$ 0.4& 15.5 $\pm$ 0.4\\ 
\textbf{42.0 $<$ log(L$_{14-195keV}$) $\leq$ 43.0}& 26.2 $\pm$ 0.5 & 20.0 $\pm$ 0.5 & 18.3 $\pm$ 0.4\\ 
\textbf{43.0 $<$ log(L$_{14-195keV}$) $\leq$ 44.2}& 25.0 $\pm$ 1.4 & 15.5 $\pm$ 0.3 & 22.7 $\pm$ 1.0\\ 
\enddata
\tablecomments{Error values represent the standard error.}
\end{deluxetable*}

\subsection{AGN Power and Line Luminosity} \label{XrayLum_vs_LineLum}

\begin{figure*}[ht]
\plotone{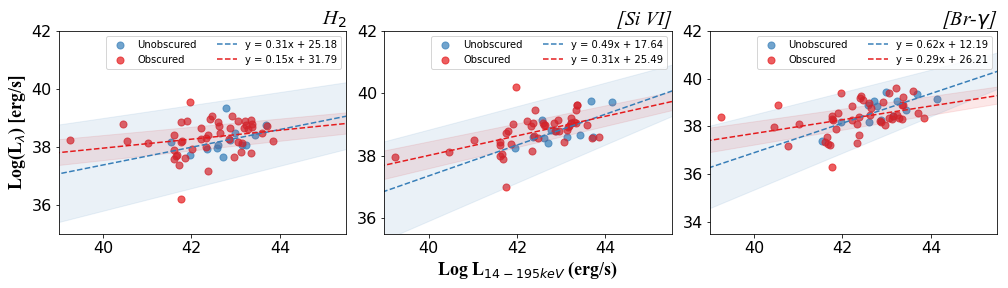}
\caption{ H$_2$ 1-0 S(1), [Si VI], and Br-$\gamma$ line luminosity within the central 200 pc aperture as a function of X-ray Luminosity for obscured and unobscured AGN. Line luminoisities were calculated from a sum of all spaxels within the central 200 pc region with a S/N $>$ 3. Dashed line represents the best fit of the linear regression and shaded region represents the 95$\%$ confidence interval. 
\label{fig13}}
\end{figure*}

We perform a linear regression to assess the relationship between emission line strength within the central 200 pc and L$_{14-195 keV}$ for the obscured and unobscured Seyfert AGN subgroups (Figure \ref{fig13}). Table \ref{Table 5} provides the resulting parameters for the linear regressions. We find no statistically significant correlation between the H$_2$ 1-0 S(1) line luminosity and the L$_{14-195keV}$ for the unobscured Seyfert AGN and a slight trend with the obscured Seyfert AGN. For [Si VI] and Br-$\gamma$ we identify a similar positive correlation between L$_{14-195 keV}$ and line luminosity for the obscured and unobscured galaxies, which suggests that regardless of obscuration these line luminosities scale with AGN power (i.e. rate of accretion onto the SMBH) and are thus driven by AGN activity. While Br-$\gamma$ emission could also be associated with non-AGN excitation mechanisms (e.g. star formation) within the central 200pc integrated spectra, the correlation with L$_{14-195keV}$ suggests a significant contribution from the AGN. Our results for [Si VI] are consistent with the positive correlation found by \cite{den2022bass} for the much larger BASS sample over larger spatial scales, however we find a flatter fit to the data. We interpret the difference in behavior of the molecular and ionized gas to result from the association of the H$_2$ 1-0 S(1) line with the relatively undisturbed gas within the rotating disk, and a lack of a strong dependence on the AGN for excitation. We do acknowledge that, over large timescales, an increased presence of hydrogen gas could eventually fuel the AGN increasing L$_{14-195keV}$, but may take significant time to feed down to the SMBH. Conversely, [Si VI] and (potentially Br-$\gamma$) are associated with high energy outflows from the AGN.  

\begin{deluxetable*}{ccccccc}
\tabletypesize{\scriptsize}
\tablewidth{0pt} 
\tablecaption{Results for the Linear regression analysis of line luminosity and the AGN log(L$_{14-195keV}$).  \label{Table 5}}
\tablehead{
\colhead{ } & \multicolumn{3}{c}{Unobscured} & \multicolumn{3}{c}{Obscured}\\
\cline{2-7}
\colhead{Emission Line} & \colhead{Slope} &
\colhead{Intercept} & \colhead{p-value} & \colhead{Slope} & \colhead{Intercept}& \colhead{p-value}
} 
\colnumbers
\startdata 
\textbf{H$_2$ 1-0 S(1)}& 0.31 $\pm$ 0.22& 25.22 $\pm$ 9.23& 0.027 & 0.15 $\pm$ 0.093& 31.79 $\pm$ 3.92 & 0.051\\ 
\textbf{[Si VI]}& 0.49 $\pm$ 0.17& 17.64 $\pm$ 7.22& 0.013 & 0.31 $\pm$ 0.082& 25.49 $\pm$ 3.50 & 0.00048\\ 
\textbf{Br-$\gamma$}& 0.62 $\pm$ 0.17& 12.19 $\pm$ 7.47& 0.0046 & 0.29 $\pm$ 0.10& 26.21 $\pm$ 4.34 & 0.0074\\ 
\enddata
\tablecomments{Error presented represents the standard error of the parameters. The p-value indicates the statistical significance of the linear trend.}
\end{deluxetable*}

\subsection{Trends of Circumnuclear Properties with Obscuration
} \label{subsec:Data Trends}

\begin{figure*}[ht]
\plotone{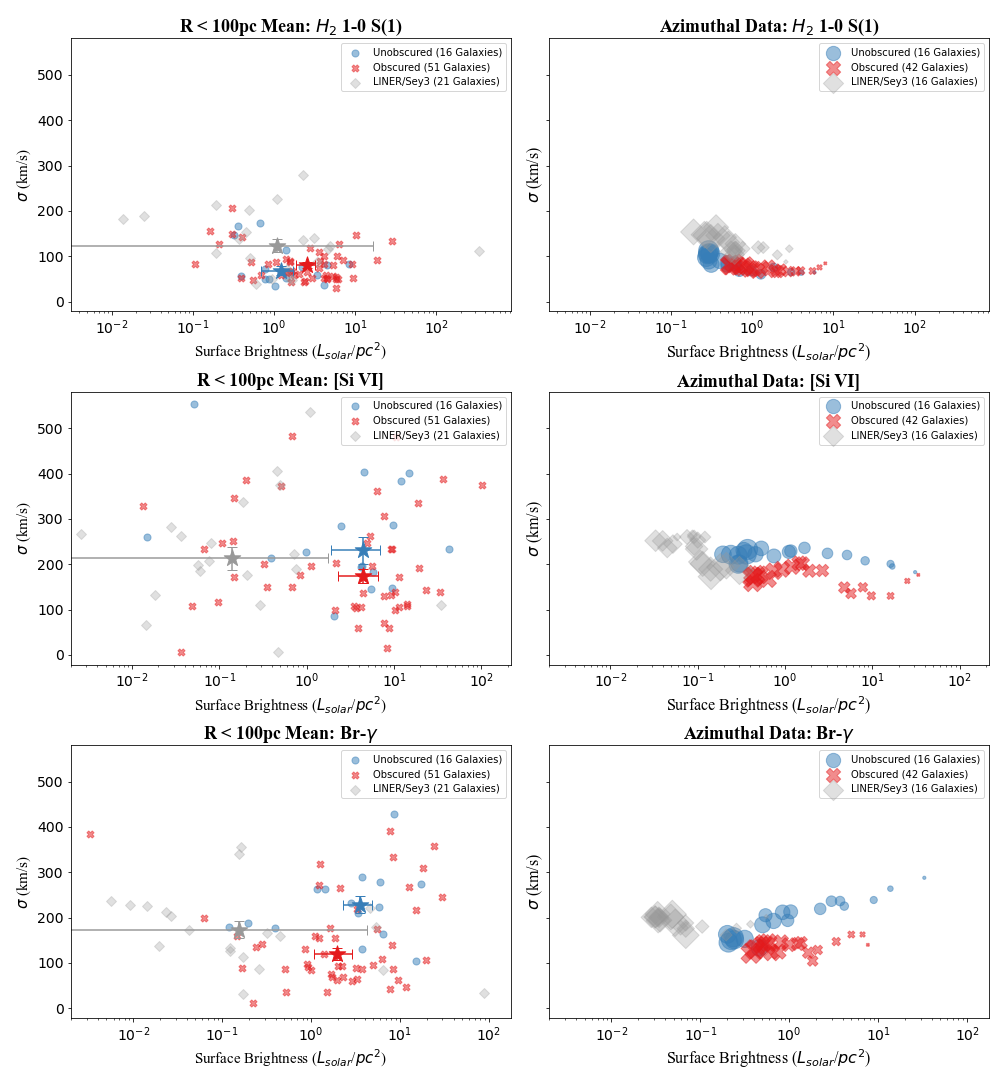}
\caption{Velocity Dispersion vs. surface brightness for H$_2$  1-0 S(1), [Si VI], and Br-$\gamma$ as denoted at the top of each panel for each AGN subgroup as indicated in the legends.  Left panels: The surface brightness and average values for velocity dispersion within the central 200 pc around the galactic nucleus for individual galaxies. Error bars represent the standard error of the distribution. The H$_2$, [Si VI], and Br-$\gamma$ plots have been truncated to display to better display the majority data.  For these emission lines, 2 (an obscured AGN and a LINER), 3 (one obscured AGN and two LINERs), and 4 (one unobscured, two obscured, and one LINER) galaxies respectively have surface brightness below what is displayed. The measured surface brightness and mean velocity dispersion for the central 200 pc aperture for each galaxy is presented in Table \ref{Table_A2_ReviewerSuggested}. Stars indicate the median values for each AGN subgroup. Right panels: The median of the azimuthal averages out to 200 pc from the galactic nucleus for each AGN subgroup.  Each point represents the median of the surface brightness and the azimuthally averaged value for of velocity dispersion for the AGN subgroup in 10 pc circular annuli. Radial distance from the galactic nucleus is inversely proportional to the point size (e.g. the largest data points represents the median values in a circular annulus at 190-200 pc and the smallest data points represent the central 10 pc).   
\label{fig9}}
\end{figure*}

Figure \ref{fig9} shows the relationship between $\sigma$ and surface brightness for the sample categorized by obscuration. Measurements for all three emission lines within the central 200 pc region for each galaxy are shown (left panels), as well as the azimuthal data out to r = 200 pc (right panels). Table \ref{Table_A2_ReviewerSuggested} presents the measured surface brightness and mean $\sigma$ for each object within the catalog. Table \ref{table 3} quantifies the subgroup median values for $\sigma$ and surface brightness (indicated in Figure \ref{fig9} as a star) along with the associated standard error for each obscuration bin. It is immediately clear that there is significant overlap in the distributions of each Seyfert sub-grouping. We do observe a slightly greater H$_2$ 1-0 S(1) median surface brightness for the central 200 pc of obscured relative to unobscured AGN. We note negligible separation between the median values of surface brightness in the central 200 pc region between unobscured and obscured Seyfert AGN for the [Si VI] or Br-$\gamma$ emission. It is possible that this increased H$_2$ 1-0 S(1) surface brightness is the result of the distinct viewing angle between obscured and unobscured Seyfert AGN assumed by the unified AGN model. There is potential that some obscuration may come from the galactic disk and therefore our measured surface brightness for obscured AGN may be supplemented by warm molecular gas from the disk.  This observation could also suggest that there is simply more warm molecular gas circumnuclear to the AGN in obscured galaxies. A mild positive correlation has been shown for cool molecular gas in previous works \citep{rosario2018llama}.

No significant difference is found in $\sigma$ for H$_2$ 1-0 S(1), however we do observe greater median $\sigma$ for unobscured AGN within the central 200 pc for [Si VI] and Br-$\gamma$. The observed difference in $\sigma$ of the [Si VI] line is notable as the difference is larger than the standard error of the measurement (Table \ref{table 3}).  Regarding Br-$\gamma$ we note that for our unobscured (Seyfert 1-1.5) galaxies, measurements within the central r $<$ 100 pc region contain some level of contamination from the broad line component. The presence of this contamination likely results in the measured values representing upper limits for average $\sigma$, which could explain the higher $\sigma$ seen in unobscured AGN, particularly at smaller radii.  

Generally, these relationships between surface brightness, $\sigma$, and obscuration are consistent for all radial distances from the nucleus for all emission lines measured (Figure \ref{fig9} right pannels, see also Figure \ref{figA1}). At all radial distances from the nucleus out to r $\approx$ 200 pc we note decreased surface brightness for H$_2$ 1-0 S(1) and increased $\sigma$ of [Si VI] and Br-$\gamma$ for unobscured relative to obscured AGN. In addition, H$_2$ 1-0 S(1) has increased $\sigma$ at greater radii from the nucleus (r $>$ 140pc) in unobscured compared to obscured AGN. The increased surface brightness could indicate an increased mass density of molecular gas in the nuclear region of obscured AGN. The observed differences in $\sigma$ could be related to the increased N$_H$ of obscured AGN shielding ionizing radiation. We acknowledge that this observation could be the result of a difference in the angle of the circumnuclear disk relative to AGN obscuration, or an artifact of the line of sight viewing angle of the AGN which results in an unobscured Seyfert classification rather than obscured (consistent with the classical unification model). 

For both unobscured and obscured AGN, the median values of $\sigma$ across the region for the H$_2$ 1-0 S(1) emission line is significantly lower than that observed for the [Si VI] and Br-$\gamma$ emission lines (Table \ref{table 3}).  We take this observation to suggest that the observed H$_2$ emission are largely disk associated whereas [Si VI] and Br-$\gamma$ are associated more closely with outflows. Similarly, median values of the surface brightness for H$_2$ 1-0 S(1) is lower than that observed for [Si VI] and Br-$\gamma$. However we anticipate that this observation may result simply from differences in the native infrared emissivity as well as the increased energy associated with outflows generating bright emission lines relative to disk associated H$_2$ emission.

\begin{deluxetable*}{ccccccc}
\tabletypesize{\scriptsize}
\tablewidth{0pt} 
\tablecaption{Median Dispersion and Median Surface Brightness.  \label{table 3}}
\tablehead{
\colhead{Type} & \multicolumn{2}{c}{H$_2$ 1-0 S(1)} & \multicolumn{2}{c}{[Si VI]} & \multicolumn{2}{c}{Br-$\gamma$} \\
\cline{1-7}
\colhead{ } & \colhead{$\sigma$} &
\colhead{Surface Brightness} & \colhead{$\sigma$} & \colhead{Surface Brightness} & \colhead{$\sigma$} & \colhead{Surface Brightness}\\
\colhead{ } & \colhead{km s$^{-1}$} &
\colhead{L$_{\odot}$ pc$^{-2}$} & \colhead{km s$^{-1}$} & \colhead{L$_{\odot}$ pc$^{-2}$} & \colhead{km s$^{-1}$} & \colhead{L$_{\odot}$ pc$^{-2}$}
} 
\colnumbers
\startdata 
\textbf{Unobscured}& 68 ± 11& 1.2 ± 0.5& 230 ± 29& 4.4 ± 2.5& 229 ± 18& 3.5 ± 1.2\\
\textbf{Obscured}& 80 ± 6& 2.5 ± 0.7& 173 ± 16& 4.4 ± 2.1& 119 ± 14& 2.0 ± 0.9\\
\enddata
\tablecomments{Median values for $\sigma$ and surface brightness for the 200 pc region of each
  Seyfert type. Error values represent the standard error of the data set. Values presented for Unobscured AGN Br-$\gamma$ represent upper limits as no correction has been applied for BLR contamination.}
\end{deluxetable*}

Somewhat analogous to surface brightness, the mass of warm molecular H$_2$ gas available to feed the AGN can be estimated from the observed flux by a method first derived by \cite{scoville1982velocity} and later applied to the circumnuclear region around AGN (e.g. \citealt{riffel2018gemini} and \citealt{schonell2019gemini}). Following this approach, the mass of total warm molecular gas is given by 
 
\begin{equation}
M_{H_{2}} = 5.0776 \times 10^{13}\cdot \left( \frac{F_{H_2\lambda2.1218}}{erg s^{-1}cm^{-2}} \right) \cdot \left( \frac{D}{Mpc} \right)^{2}\cdot M_{\odot} \label{H2_Mass}
\end{equation}

\begin{figure*}[ht]
\plotone{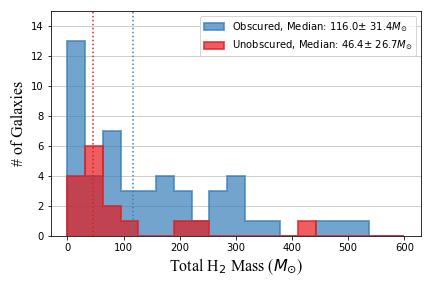}
\caption{Histogram of the central 200 pc total molecular hydrogen mass for the obscured and unobscured subgroups. One galaxy, the obscured galaxy NGC 1068, yielded a molecular hydrogen mass of $\approx$ 1450 M$_{\odot}$ and is not included in this histogram for display purposes. 
\label{Obscuration_TotalMass_Histogram}}
\end{figure*}

Equation \ref{H2_Mass} \citep{schonell2019gemini}, assumes the molecular gas to be thermalized with a vibrational temperature of 2000 K. Figure \ref{Obscuration_TotalMass_Histogram} depicts a histogram for the total warm molecular H$_2$ mass for the central 200 pc nuclear region of each galaxy within the sample. We report estimates for the median of warm H$_2$ mass within the central 200 pc region to be 46.4 $\pm$ 26.7 $M_{\odot}$ for unobscured AGN and 116.0 $\pm$ 31.4 $M_{\odot}$ for obscured AGN, suggesting that obscured AGN have roughly twice as much molecular gas within the central 200pc. 

\subsection{Trends of Circumnuclear Properties with X-ray Luminosity} \label{subsec:Data Trends Xray}

\begin{figure*}[ht]
\plotone{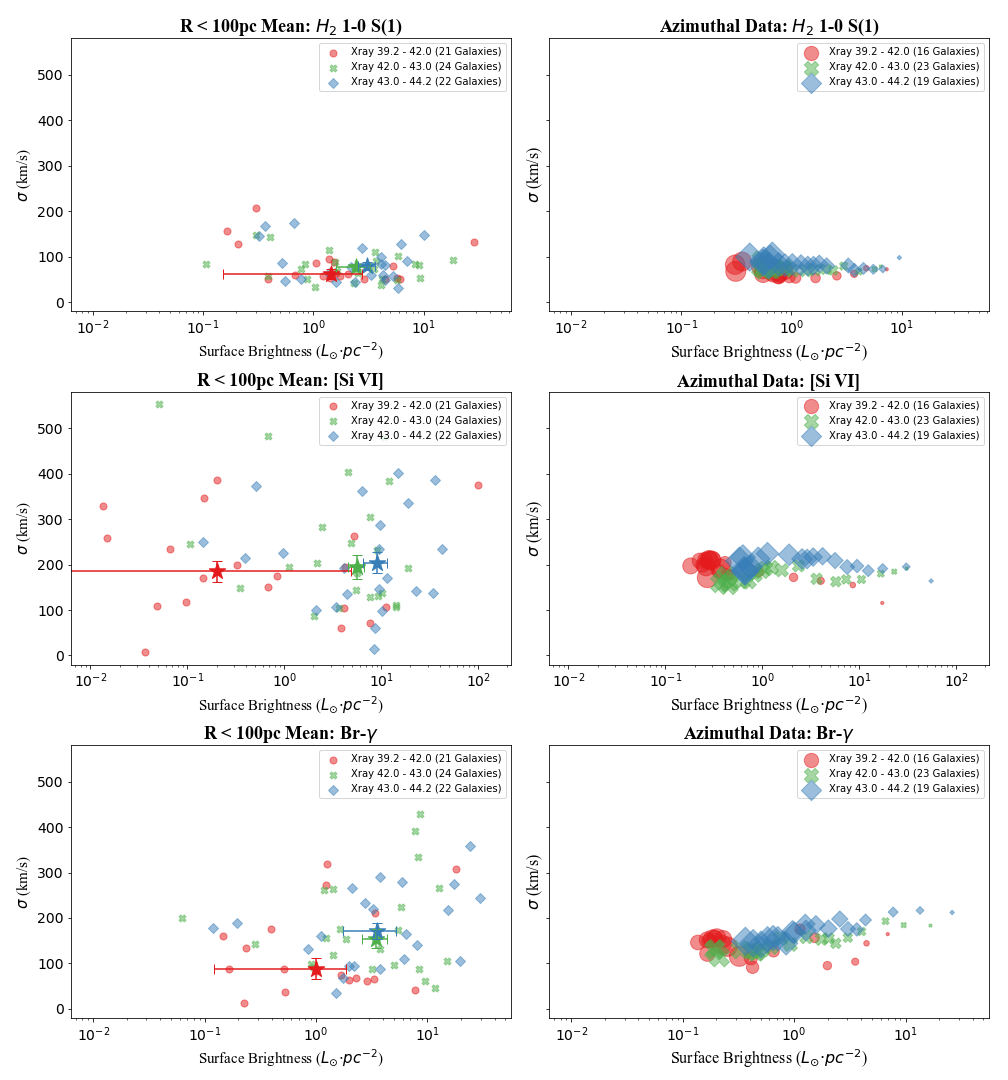}
\caption{ Velocity dispersion vs. surface brightness for H$_2$  1-0 S(1), [Si VI], and Br-$\gamma$ as denoted at the top of each panel as a function of X-ray luminosity as indicated in the legends.  Left panels: The average values of $\sigma$ plotted against the surface brightness within the central 200 pc around the galactic nucleus for individual galaxies. Error bars represent the standard error of the distribution. Stars indicate the median values for each luminosity subgroup. The H$_2$ 0-1 S(1), [Si VI], and Br-$\gamma$ plots have been truncated to better display the majority of the data.  One galaxy, from the lowest X-ray luminosity bin exhibited a surface brightness below what is displayed for H$_2$ 0-1 S(1) and [Si VI]. Four galaxies, three from the lowest X-ry luminosity bin, and one from the 42.0-43.0 L$_{14-195 keV}$ bin, exhibited a surface brightness below what is displayed for Br-$\gamma$. The measured surface brightness and mean velocity dispersion for the central 200 pc aperture for each galaxy is presented in Table \ref{Table_A2_ReviewerSuggested}.  Right panels: The median of the azimuthal averages out to 200 pc from the galactic nucleus for each X-ray luminosity subgroup.  Each point represents the median of the surface brightness and the median average velocity dispersion for the AGN subgroup in 10 pc circular annuli. Radial distance from the galactic nucleus is inversely proportional to the point size (e.g. the largest data points represents the median values in a circular annulus at 190-200 pc and the smallest data points represent the central 10 pc).   
\label{fig10}}
\end{figure*}

Figure \ref{fig10} depicts the relationship between $\sigma$ and surface brightness in all galaxies categorized by the L$_{14-195keV}$. Table \ref{Table 4} quantifies the median values for $\sigma$ and surface brightness within the central 200 pc region  for each emission line (indicated in Figure \ref{fig10} as a star), and also provides the associated standard errors. We again note significant overlap in the distributions of each L$_{14-195keV}$ subgroup for surface brightness and average $\sigma$ within the central 200 pc (Figure \ref{fig10} left panels). We report no strong correlation between $\sigma$ and L$_{14-195keV}$, although we do observe the median $\sigma$ for Br-$\gamma$ to be notably lower for the lowest L$_{14-195kev}$ bin. Median values of surface brightness do show some positive correlation with L$_{14-195keV}$ for all three emission lines, however these differences fall within the standard error of the data. While the standard deviation of the datasets is large, this observation could be symptomatic of the overall reduced local intensity of ionising radiation energy within the circumnuclear region for the lower L$_{14-195keV}$ bins. 

These trends are again generally consistent for all radial distances from the nucleus (Figure \ref{fig10} and \ref{figA3}). The H$_2$ 1-0 S(1) emission line presents the most consistent correlation with L$_{14-195 keV}$ at each radial bin, with increasing surface brightness and a slight increase in $\sigma$ with L$_{14-195 keV}$. While differences in surface brightness are generally within the standard error of the data, $\sigma$ for the range of $\approx$ 50-150 pc presents some separation outside of the standard error for between the highest and lowest L$_{14-195 keV}$ bin (Figure \ref{figA3}). [Si VI] displays no significant difference in measured $\sigma$ for each X-ray luminosity bin. Surface brightness measurements for [Si VI] do show some slight correlation with L$_{14-195keV}$. This behavior is largely mimicked in Br-$\gamma$ with the exception of some difference in measured $\sigma$, with the lowest X-ray luminosity bin yielding lower values of $\sigma$ for measurements within the r $<$ 100 pc region. We take this lower value in sigma to be a symptom of a greater prevalence of broad line contamination within the central region for more luminous AGN. 

For all X-ray luminosity bins we observe increased values for $\sigma$ and surface brightness the [Si VI] and Br-$\gamma$ emission lines over the H$_2$ 0-1 S(1) emission. This observation is apparently indifferent to parsing data by obscuration or L$_{14-195 keV}$. Again, we reiterate, that this difference could be the result of the native emissivity of each emission lines and/or resultant of [Si VI] and Br-$\gamma$ being primarily associated with higher energy environments such as outflows. The latter would again reinforce that the bulk of molecular hydrogen gas is entrained in more uniform disk rotation and is excited by a different suite of  mechanisms than [Si VI] and Br-$\gamma$ \citep{riffel2020ionized}.  

\begin{deluxetable*}{ccccccc}
\tabletypesize{\scriptsize}
\tablewidth{0pt} 
\tablecaption{Median Dispersion and Median Surface Brightness for X-ray Luminosity bins.  \label{Table 4}}
\tablehead{
\colhead{log(L$_{14-195keV}$)} & \multicolumn{2}{c}{H$_2$ 1-0 S(1)} & \multicolumn{2}{c}{[Si VI]} & \multicolumn{2}{c}{Br-$\gamma$} \\
\cline{1-7}
\colhead{ } & \colhead{$\sigma$} &
\colhead{Surface Brightness} & \colhead{$\sigma$} & \colhead{Surface Brightness} & \colhead{$\sigma$} & \colhead{Surface Brightness}\\
\colhead{ } & \colhead{km s$^{-1}$} &
\colhead{L$_{\odot}$ pc$^{-2}$} & \colhead{km s$^{-1}$} & \colhead{L$_{\odot}$ pc$^{-2}$} & \colhead{km s$^{-1}$} & \colhead{L$_{\odot}$ pc$^{-2}$}
} 
\colnumbers
\startdata 
39.2 - 42.0& 62 ± 11& 1.5 ± 1.3& 185 ± 23& 0.2 ± 4.6& 88 ± 23& 1.0 ± 0.9\\
42.0 - 43.0& 78 ± 6& 2.4 ± 0.8& 195 ± 27& 5.6 ± 1.0& 154 ± 20& 3.5 ± 0.9\\
43.0 - 44.2& 79 ± 9& 3.1 ± 0.5& 204 ± 23& 9.0 ± 2.5& 172 ± 18& 3.5 ± 1.8\\
\enddata
\tablecomments{Mean values for velocity dispersion and surface brightness for the central 200 pc region of each X-ray Luminosity bin. Error presented represents the standard error. Values presented for Unobscured AGN Br-$\gamma$ represent upper limits as no correction has been applied for BLR contamination.}
\end{deluxetable*}

\begin{figure*}[ht]
\plotone{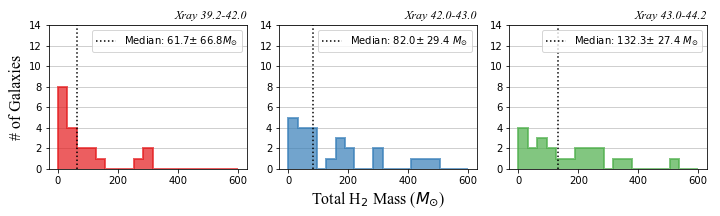}
\caption{Histogram of the central 200 pc total Molecular Hydrogen mass for each X-ray luminosity subgroup. One galaxy,NGC 1068 within the lowest X-ray luminosity bin, yielded a molecular hydrogen mass of $\approx$ 1450 M$_{\odot}$ and is not included in this histogram for display purposes. 
\label{XrayLum_TotalMass_Histogram}}
\end{figure*}

Again, following the method derived by \cite{scoville1982velocity}, we estimate the mass of warm molecular H$_2$ gas within the central 200 pc circumnuclear region for each L$_{14-195keV}$ regime. Figure \ref{XrayLum_TotalMass_Histogram} displays a histogram of the total molecular hydrogen mass calculated for each of the L$_{14-195keV}$ subgroups. We warm H$_2$ mass from lowest to highest L$_{14-195 keV}$ group to be 61.7 $\pm$ 66.8 $M_{\odot}$, 82.0 $\pm$ 29.4 $M_{\odot}$, and 132.3 $\pm$ 27.4 $M_{\odot}$ respectively. We report a weak positive correlation between the L$_{14-195keV}$ and the amount of warm molecular gas around the AGN. 

 \subsection{Radial Dependence of Parameters} \label{subsec:Azimuthal Average}

Our measurements are in agreement with previous studies \citep{hicks2016keck}, in which no statistically significant radial dependence of $\sigma$ was found for either unobscured or obscured Seyfert AGN for the H$_2$ 1-0 S(1), [Si VI], and Br-$\gamma$ emission lines within the circumnuclear region (outside of increases which can be attributed to broad line contamination; Figure \ref{fig9}, \ref{figA1}).  With regards to L$_{14-195 keV}$, we again observe no statistically significant radial dependence of $\sigma$ (Figure \ref{fig10}, \ref{figA3}).  We take these observations to suggest that AGN radiative feedback itself is not a significant contributor to the velocity dispersion $\sigma$, but rather $\sigma$ may be governed by other mechanisms such as turbulence which could be stimulated by outflows. Such an  interpretation is consistent with the increased value of $\sigma$ for [Si VI] (associated with outflows) relative to the H$_2$ 1-0 S(1) emission (associated with the more orderly rotational disk) \citep{riffel2020ionized}.

Surface brightness on the other hand, decreases precipitously with growing distance from the nucleus. For all subgroups considered, the median surface brightness for each of the measured emission lines decreases by at least 80\% within the central 100 pc. The observed decrease we interpret to result from both the geometrical dilution of AGN ionizing radiation and of a non-uniform mass density gradient. As radial distance increases, we expect the AGN radiative energy density for excitation decreases which would result in decreased emission. Similarly, we may expect the mass density to decrease with radial distance from the nucleus, thereby resulting in less gas to excite and therefore decreased surface brightness. 

\section{Discussion} \label{sec:Discussion}

This analysis has presented a population-wide look at the behavior of the molecular and ionized gas traced in the NIR in the circumnuclear region around AGN. We have examined the behavior of $\sigma$ and surface brightness in relation to different fundamental AGN characteristics (obscuration and X-ray luminosity) as well as radial distance from the nucleus. Due to the significant scatter of the measured circumnuclear characteristics of all three gas tracers in this sample it is difficult to assess the root cause of some observed differences. While this prevents drawing firm conclusions, this study does point toward differences in the AGN’s influence over the behaviour of the molecular and ionized gas within the circumnuclear region.  

As a whole we consider the impacts of AGN radiative feedback and potential outflows to help constrain how the AGN is influencing its local environment. From our analysis we confirm the findings of previous studies in which ionized gas tends to be more centrally located relative to molecular gas \citep{schonell2019gemini, riffel2021agnifs}. We take this to suggest that the AGN enhancement of warm molecular hydrogen extends farther than the ionizing radiation which drives the higher energy [Si VI] and Br-$\gamma$ emission. That is to suggest that the molecular gas may be thermalized by radiative and collisional processes that extend into the disk, whereas the ionization energy needed to produce [Si VI] and Br-$\gamma$ emission resides closer to the AGN. We also anticipate that the bulk of the molecular hydrogen is more strongly associated with the rotational disk the AGN compared to the ionized gases. This is supported by our observation of the generally lower $\sigma$ observed for molecular hydrogen relative to [Si VI] and Br-$\gamma$, implying more orderly kinematics of molecular hydrogen relative to ionized gases. As such, the presence of H$_2$ is likely associated primarily with AGN feeding rather than feedback as has been highlighted in previous works \citep{storchi2019observational}.

Regarding AGN radiative feedback, our observations of a slight correlation between L$_{14-195keV}$ and observed surface brightness could lead to the conclusion that radiative feedback is an influential source of excitation energy for local gases of the circumnuclear region. However, our preference for this interpretation is marred by the potential for differences in mass density gradients. Such differences could result in the observed correlation as a higher mass density could imply an increased accretion rate and therefore increased L$_{14-195keV}$. Such an interpretation would be consistent with the weak positive correlation we report for surface mass density within the nuclear region (Section \ref{subsec:Data Trends Xray}). On the other hand, $\sigma$, with its lack of
 correlation with L$_{14-195keV}$ and its apparent indifference to radial distance, suggest that radiative excitation does not influence this parameter. As an alternative, we suggest that the kinematics of the local environment drive $\sigma$. We take this as  further evidence that molecular hydrogen emission are disk associated while [Si VI] and Br-$\gamma$ are associated with more complex kinematic environments, such as outflows. This is also supported by our observations of line luminosity scaling of [Si VI] and Br-$\gamma$ with L$_{14-195keV}$ and, in contrast to a lack a similar correlation with  H$_2$ 1-0 S(1) emission. Overall, our interpretation is that the AGN does provide ionizing radiation which warms and excites the gaseous material within the central 400 pc region, however it does not seem to drive the $\sigma$. Rather we suspect $\sigma$ is more dominantly controlled by other mechanisms such as the impacts of outflows on the local environment and turbulence of gaseous material. 

Perhaps the more significant observations in this study were those between obscured and uobscsured AGN. Utilizing this larger sample we observed, on average, increased surface brightness of H$_2$ 1-0 S(1) in the circumnuclear region of obscured AGN relative to unobscured AGN. We take this to suggest that obscured AGN have a higher mass density of molecular gas in circiumnuclear region relative to unobscured AGN, an interpretation that is consistent with the surface mass density of warm molecular hydrogen calculated in section \ref{subsec:Data Trends}. This result is consistent with previous studies which suggested galaxies which house an obscured AGN contain a denser nuclear environment \citep{rosario2018llama, almeida2011testing}. Another interesting observation in this study is the increased $\sigma$ for unobscured [Si VI] emission relative to obscured AGN. Taking [Si VI] to be primarily associated with outflows, we are tempted to speculate that outflows are intrinsically different in obscured AGN. However, we again reiterate that the scatter of the dataset is significant and observations of $\sigma$ binned by obscuration could be subject to a selection bias governed by the the line of sight viewing angle. It is possible that the edge on orientation of some galaxies resulting in the obscured categorization of the object may have some influence on our observations of outflows and the measured dispersion. Future studies in which circumnuclear disk orientations are properly taking into account may reveal a better understanding of this observed difference.

Our analysis is focused on the innermost regions (r $<$ 200 pc) around the AGN. Investigations of ionized gas emission, such as [O III], in similar AGN samples have indicated the presence of outflows and disturbed gas out to kpc scales (e.g. \citealt{harrison2014kiloparsec, polack2024determining, revalski2021quantifying, kang2018unraveling}). The impact of the AGN and behaviour of the multiphase outflow can be expected to evolve beyond the scales probed in this study. For example, the sigma of [O III] has been shown to decrease with radius out to kpc scales \citep{kang2018unraveling, karouzos2016unraveling}. As such, to obtaining a more full understanding of the connection between the AGN and the host galaxy, will require careful analysis of impacts in both the immediate vicinity of the AGN as well as at broader scales.

While our observations are underscored by the impressive variation in the surface brightness and $\sigma$ across the Seyfert subgroups considered, we find the this to be both interesting and informative. Ultimately, we take this as an indicator that the circumnuclear region is wildly diverse across the Seyfert galaxy population, both in terms of kinematics and excitation, implying a varied set of processes are important within the region rather than simply AGN radiative feedback. In fact, we take the large scatter in measured properties across the sample to be a primary takeaway of this analysis, implying that there may be other underlying fundamental relationships at play which require investigation. This highlights a need for further study across a variety of AGN to tease out and better constrain the fundamental processes governing AGN feeding and feedback. Likely future efforts will require a combination of work focusing on individual galaxies and large scale studies. Detailed analysis on individual galaxies will be critical to identify in the range of mechanisms driving AGN feedback, and large scale studies will be act to assess the relative importance of these mechanisms in influencing the AGN-host galaxy relationship. With the advent of data made available via the LUNIS-AGN catalog and newer observatories such as JWST collecting complementary high resolution spatial data, we are optimistic at the prospect of future efforts to study and understand the behavior of AGN.

 \section{Summary} \label{sec:Interpretations}

We have generated the largest sample set of archival NIR IFU data for nearby (z $\le$ 0.02) Seyfert type AGN ever aggregated. Further, we have generated a suite of data products which will provide continued benefit and stimulate further AGN research. In addition, we took a first look at utilizing this larger sample to investigate AGN properties of gaseous material within the central 400 pc region around the galactic nucleus. Utilizing 2D line maps for the H$_2$  1-0 S(1), [Si VI], and Br-$\gamma$ emission lines we have analyzed and compared the mean $\sigma$ and surface brightness within the circumnuclear region and considered these in the context of Seyfert type (obscuration) as well as the L$_{14-125 keV}$.  From our analysis of the catalog sample, we report the following observations: 

\begin{itemize}
    \item Consistent with previous studies, the distribution of the [Si VI] and Br-$\gamma$ emission is more centrally concentrated than molecular hydrogen emission \citep{schonell2019gemini, riffel2021agnifs}.
    \item Consistent with previous works \citep{almeida2011testing, rosario2018llama}, we find an increased mass of warm molecular hydrogen in the central 200 pc region of obscured (Seyfert 2) relative to unobscured (Seyfert 1) galaxies.
    \item We measure $\sigma$ for H$_2$  1-0 S(1) to be roughly half that measured for [Si VI] and Br-$\gamma$ emission for all data subgroupings.  This supports the expectation that [Si VI] and Br-$\gamma$ emission are likely associated with kinematically different environments such as outflows. While the circumnuclear molecular hydrogen may also trace outflows in some cases, it is also associated with rotation within a circumnuclear disk and inflows \citep{riffel2020ionized}.  
    \item  $[$Si VI] and Br-$\gamma$ line luminosity distributions appear to scale with the AGN L$_{14-195keV}$ whereas the H$_2$ 1-0 S(1) line luminosity does not. This scaling is consistent with results presented by \cite{den2022bass}.     
    \item We report no radial dependence of $\sigma$ for either unobscured and obscured Seyfert Galaxies for the H$_2$ 1-0 S(1),  [Si VI], or Br-$\gamma$.  This result is consistent with previously studies \citep{hicks2016keck}.  However we do report the observation of increased velocity dispersion for unobscured AGN relative to obscured AGN. 
    \item We report no statistically significant difference in surface brightness between unobscured and obscured Seyfert AGN for [Si VI] and Br-$\gamma$. Increased surface brightness is observed for obscured  relative to unobscured AGN for the H$_2$ 1-0 S(1) emission line within the central 400 pc region. 
    \item Our data suggests surface brightness is radially dependent and decreases precipitously with distance from the nucleus. For all data subgroups and measured emission lines, we observe the median surface brightness to  decrease by at least 80\% within the central 100 pc.
    \item Surface brightness displays a slight correlation with measured L$_{14-195 keV}$ X-ray luminosity for any emission line measured. 

\end{itemize}

While this work highlights potentially meaningful differences in the circumnulear regions of local Seyfert galaxies with AGN obscuration and X-ray luminosity, it also demonstrates the significant variation in circumnuclear properties across the sample in both molecular and ionized gas. Continued efforts utilizing archival data, such as those presented in this catalog, paired with new high resolution observations (e.g. JWST) will be necessary to more fully constrain the nature of AGN behavior and connection to the evolution of galaxies.

\section*{Acknowledgments}

\begin{acknowledgments}
We thank the referee for their useful comments which have improved the manuscript. This material is based upon work supported by the National Science Foundation under Grant No. 1911242. This work was also supported by the 2022 Alaska Space Grant. This research has made use of the Keck Observatory Archive (KOA), which is operated by the W. M. Keck Observatory and the NASA Exoplanet Science Institute (NExScI), under contract with the National Aeronautics and Space Administration.  David J. Rosario  acknowledges the support of grant ST/X001105/1 from the Science and Technology Facilities Council (STFC) of the UK. This research has made use of the services of the ESO Science Archive Facility. This research has made use of the NASA/IPAC Extragalactic Database (NED), which is operated by the Jet Propulsion Laboratory, California Institute of Technology, under contract with the National Aeronautics and Space Administration. We would also like to thank those who contributed and provided datasets as well as invaluable feedback and suggestions and comments which were vital in the editing and production of this manuscript. This research utilized the unique and powerful tools included within Astropy, a community developed python package \citep{robitaille2013astropy}. Francisco Muller-Sanchez acknowledges support from NASA through ADAP award 80NSSC19K1096. 
\end{acknowledgments}

%

\vspace{5mm}
\facilities{VLT (SINFONI), Keck (OSIRIS)}


\software{astropy \citep{2013A&A...558A..33A,2018AJ....156..123A},  LINEFIT  \citep{davies2007method}
          }



\appendix
\renewcommand{\thefigure}{A\arabic{figure}}
\setcounter{figure}{0}

\renewcommand{\thetable}{A\arabic{table}}
\setcounter{table}{0}  

\section{Appendix Information}

This appendix is intended to provide additional data and supplemental data visualization for data presented in this manuscript.  This appendix presents data tables pertaining to general flux measurements, [Si VI] elliptical isophote fitting, and plots for the azimuthal average of both $\sigma$ and surface brightness for data binned by Seyfert type as well as the L$_{14-195keV}$ X-ray luminosity. 

\subsection{Measurements of the central 200 pc Region}\label{subsec:A1}
Here (Table \ref{Table_A1}) we include the data for the integrated flux measurements for the central 200 pc for H$_2$ 1-0 S(1), [Si VI], and Br-$\gamma$ emission. In Addition, Table \ref{Table_A2_ReviewerSuggested} provides surface brightness and mean dispersion measurments within the central 200 pc region. 

\begin{deluxetable*}{ccccccc}
\tabletypesize{\scriptsize}
\tablewidth{0pt} 
\tablecaption{H$_2$ 1-0 S(1), [Si VI], and Br-$\gamma$ line luminosity and half-width-half max (HWHM) over the central 200 pc region for each galaxy within the catalog sample.  \label{Table_A1}}
\tablehead{
\colhead{ } & \multicolumn{3}{c}{log[L$_{\lambda}$] Central 200 pc} &\multicolumn{3}{c}{HWHM} \\ 
\colhead{ } & \multicolumn{3}{c}{erg s$^{-1}$} &\multicolumn{3}{c}{pc} \\
\cline{2-7}
\colhead{Source Name} & \colhead{H$_2$ 1-0 S(1)} &\colhead{[Si VI]} &\colhead{Br-$\gamma$} & \colhead{H$_2$ 1-0 S(1)} &\colhead{[Si VI]} &\colhead{Br-$\gamma$}
} 
\colnumbers
\startdata 
Circinus        & 36.2             & 37.01            & 36.32            & 10  & 10  & 10  \\
ESO 137-G034    & 38.74            & 38.98            & 38.47            & 40  & 20  & 20  \\
ESO 428-G14     & 38.64            & 39.1             & 38.85            & 20  & 10  & 10  \\
ESO 548-81      & 38.06            & 39.04            & 39.07            & 10  & 20  & 20  \\
IC 1459         & 37.7             & 38.62            & --               & 20  & 30  & --  \\
IC 2560         & 38.43            & 39.06            & 38.4             & 30  & 20  & 20  \\
IC 4296         & 37.74            & 37.3             & 36.7             & 30  & 100 & 70  \\
IC 4329a        & --               & 39.71            & 39.14            & 10  & 10  & 10  \\
IC 5063         & 38.14            & 38.92            & 38.44            & 20  & 10  & 10  \\
IRAS 01475-0740 & 38.96            & 39.47            & 39.26            & 20  & 20  & 10  \\
M 87            & 37.79            & \textless{}38.38 & --               & 20  & 10  & 10  \\
MCG -05-23-016  & 38.47            & 39               & 39.49            & 10  & 20  & 20  \\
MCG -06-30-015  & 37.64            & 38.41            & 39.06            & 20  & 10  & 10  \\
MRK 1066        & 39.07            & 38.57            & 39.09            & 30  & 10  & 30  \\
MRK 1210        & 38.79            & 39.63            & 39.23            & 30  & 20  & 20  \\
MRK 573         & 37.8             & 39.04            & 38.33            & 20  & 30  & 30  \\
MRK 766         & 38.48            & --               & 39.43            & 10  & 10  & 20  \\
NGC 1052        & 38.45            & 38.06            & \textless{}37.22 & 10  & 60  & 10  \\
NGC 1068        & 39.54            & 40.21            & 39.4             & 10  & 10  & 20  \\
NGC 1097        & 38.32            & 38.21            & \textless{}37.44 & 10  & 20  & 10  \\
NGC 1194        & 37.84            & 38.83            & 38.31            & 30  & 10  & 10  \\
NGC 1320        & 37.99            & 38.95            & 38.11            & 20  & 20  & 20  \\
NGC 1365        & --               & 38.65            & 39.19            & 60  & 20  & 10  \\
NGC 1386        & 38.21            & 38.8             & 38.26            & 20  & 20  & 20  \\
NGC 1566        & 38.16            & --               & 37.38            & 10  & 10  & 10  \\
NGC 1614        & 38.18            & --               & 39.39            & 160 & 10  & 110 \\
NGC 1667        & 38.8             & 38.11            & 37.97            & 40  & 20  & 20  \\
NGC 1672        & 37.72            & 37.91            & 37.28            & 20  & --  & 40  \\
NGC 1808        & 38.18            &  --              & 38.45            & 20  & 30  & 10  \\
NGC 2110        & 38.7             & 38.56            & 38.55            & 20  & 10  & 20  \\
NGC 253         & 38.21            & --               & 38.91            & 20  & 30  & 10  \\
\enddata
\tablecomments{This table has been truncated for print. The full machine readable data table is available online. Luminosity values reported as "--" indicate an instance where Linefit was unable to achieve a reliable measurement with the given parameters. HWHM values reported as "--" indicate instances where the HWHM was not able to be reliably measured.}
\end{deluxetable*}

\begin{deluxetable*}{cccccccc}
\tabletypesize{\scriptsize}
\tablewidth{0pt} 
\tablecaption{Measured surface brightness and average $\sigma$ within the central 200 pc nuclear aperture for H$_2$ 1-0 S(1), [Si VI], and Br-$\gamma$ for each galaxy within the catalog sample.  \label{Table_A2_ReviewerSuggested}}
\tablehead{
\colhead{Galaxy} & \multicolumn{2}{c}{H$_2$ 1-0 S(1)} & \multicolumn{2}{c}{[Si VI]} & \multicolumn{2}{c}{Br-$\gamma$} & \colhead{} \\
\cline{1-8}
\colhead{ } & \colhead{$\sigma$} &
\colhead{Surface Brightness} & \colhead{$\sigma$} & \colhead{Surface Brightness} & \colhead{$\sigma$} & \colhead{Surface Brightness} & \colhead{Warm H$_2$ Mass} \\
\colhead{ } & \colhead{km s$^{-1}$} &
\colhead{L$_{\odot}$ pc$^{-2}$} & \colhead{km s$^{-1}$} & \colhead{L$_{\odot}$ pc$^{-2}$} & \colhead{km s$^{-1}$} & \colhead{L$_{\odot}$ pc$^{-2}$} & \colhead{M$_{\odot}$} 
} 
\colnumbers
\startdata 
           Circinus &        57.7$\pm$0.1 &              1.8$\pm$0.1 &         106.1$\pm$0.1 &                  11.3$\pm$1.1 &          69.0$\pm$0.1 &                   2.3$\pm$0.2 &              0.7$\pm$2.5e-07 \\
    ESO  137-G  034 &       109.6$\pm$2.6 &              3.6$\pm$1.6 &         195.1$\pm$5.0 &                   5.6$\pm$3.6 &         154.5$\pm$3.8 &                   1.9$\pm$1.1 &            183.3$\pm$1.2e-02 \\
           ESO  428 &        73.5$\pm$1.2 &              3.6$\pm$2.9 &         137.7$\pm$2.5 &                   10.1$\pm$10 &         174.2$\pm$2.7 &                   5.5$\pm$4.5 &            182.4$\pm$4.6e-03 \\
           ESO  548 &        50.8$\pm$5.6 &                0.8$\pm$1 &        287.5$\pm$12.7 &                   9.8$\pm$8.6 &        278.6$\pm$16.4 &                   5.9$\pm$6.9 &             38.9$\pm$2.7e-02 \\
           IC  1459 &       139.0$\pm$5.0 &              0.4$\pm$0.6 &        405.6$\pm$13.9 &                     0.5$\pm$1 &         235.9$\pm$5.2 &                 5.6e-03$\pm$0 &             18.8$\pm$1.6e-03 \\
           IC  2560 &        50.9$\pm$1.3 &              4.2$\pm$3.3 &         111.3$\pm$2.6 &                 14.2$\pm$16.2 &          88.3$\pm$2.3 &                   3.2$\pm$3.4 &            212.5$\pm$2.9e-02 \\
           IC  4296 &      225.6$\pm$20.2 &              1.1$\pm$0.6 &        189.7$\pm$15.1 &                   0.7$\pm$0.3 &          30.5$\pm$3.8 &                   0.2$\pm$0.1 &                 56.0$\pm$0.2 \\
          IC  4329a &       173.1$\pm$9.4 &              0.7$\pm$1.6 &        401.4$\pm$20.2 &                 14.9$\pm$27.5 &        233.2$\pm$13.8 &                   2.8$\pm$8.1 &             33.9$\pm$2.0e-02 \\
           IC  5063 &       146.7$\pm$3.8 &             10.1$\pm$6.3 &         142.6$\pm$4.3 &                   22.9$\pm$28 &         139.7$\pm$4.4 &                   8.1$\pm$9.1 &            515.9$\pm$6.0e-02 \\
              M  87 &       212.0$\pm$5.1 &              0.2$\pm$0.3 &         176.3$\pm$3.7 &                   0.2$\pm$0.5 &         212.7$\pm$4.1 &                     0$\pm$0.1 &              9.7$\pm$1.9e-04 \\
MCG-05-23-016 &        44.6$\pm$1.3 &              2.4$\pm$2.3 &          14.6$\pm$0.5 &                   8.3$\pm$9.3 &         357.5$\pm$7.6 &                 24.3$\pm$31.2 &            122.6$\pm$1.2e-02 \\
MCG-06-30-015 &        56.9$\pm$7.6 &              0.4$\pm$0.2 &          86.9$\pm$3.3 &                     2$\pm$1.7 &        429.2$\pm$16.7 &                   8.6$\pm$7.9 &             19.8$\pm$8.0e-03 \\
          MRK  1066 &        52.6$\pm$1.6 &                9.3$\pm$5 &         203.1$\pm$9.4 &                   2.2$\pm$5.3 &          61.8$\pm$1.9 &                   9.5$\pm$6.3 &            474.6$\pm$7.9e-02 \\
          MRK  1210 &        30.0$\pm$1.3 &              5.9$\pm$4.3 &        387.2$\pm$12.3 &                 36.1$\pm$37.4 &         217.2$\pm$7.4 &                 15.1$\pm$16.9 &            267.4$\pm$3.9e-02 \\
           MRK  573 &        86.9$\pm$7.2 &              0.5$\pm$0.7 &          98.5$\pm$4.2 &                  10.2$\pm$6.7 &          93.6$\pm$6.2 &                     2$\pm$1.5 &             24.5$\pm$1.2e-02 \\
           MRK  766 &        37.3$\pm$4.9 &              4.1$\pm$4.1 &         182.0$\pm$9.1 &                  5.7$\pm$12.8 &         104.4$\pm$6.7 &                 15.1$\pm$18.3 &            208.5$\pm$8.9e-02 \\
          NGC  1052 &       135.4$\pm$2.3 &              2.3$\pm$2.5 &         375.8$\pm$8.5 &                   0.5$\pm$0.5 &         166.8$\pm$3.8 &                   0.3$\pm$0.9 &            116.7$\pm$2.7e-03 \\
          NGC  1068 &       132.5$\pm$2.0 &            28.6$\pm$28.5 &         374.2$\pm$6.0 &               100.4$\pm$224.2 &         308.8$\pm$4.3 &                 18.2$\pm$33.4 &           1452.6$\pm$1.1e-02 \\
          NGC  1097 &        46.6$\pm$0.7 &              1.7$\pm$1.3 &         197.8$\pm$5.4 &                   0.1$\pm$0.2 &         173.7$\pm$3.7 &                     0$\pm$0.1 &             85.5$\pm$8.6e-04 \\
          NGC  1194 &        46.8$\pm$5.5 &              0.6$\pm$0.6 &         136.1$\pm$8.9 &                   4.4$\pm$6.8 &          35.6$\pm$3.5 &                   1.5$\pm$2.1 &             28.1$\pm$1.4e-02 \\
          NGC  1320 &        83.4$\pm$5.4 &              0.8$\pm$1.1 &         129.7$\pm$3.0 &                  7.7$\pm$10.3 &          92.1$\pm$5.2 &                   0.9$\pm$1.3 &             40.9$\pm$7.2e-03 \\
          NGC  1365 &        83.6$\pm$2.6 &              0.1$\pm$0.2 &         103.5$\pm$1.7 &                   3.7$\pm$4.5 &         266.6$\pm$2.9 &                 12.7$\pm$28.1 &              5.3$\pm$1.1e-04 \\
          NGC  1386 &        94.7$\pm$1.3 &              1.4$\pm$1.4 &         263.2$\pm$4.0 &                   5.3$\pm$8.9 &         271.6$\pm$4.7 &                   1.2$\pm$1.7 &             70.6$\pm$7.5e-04 \\
          NGC  1566 &        52.0$\pm$0.8 &              1.4$\pm$1.3 &         259.5$\pm$2.7 &                     0$\pm$0.1 &         176.2$\pm$2.6 &                   0.4$\pm$1.2 &             63.0$\pm$3.2e-04 \\
          NGC  1614 &        43.3$\pm$6.0 &              1.6$\pm$0.7 &        250.7$\pm$30.8 &                   0.1$\pm$0.3 &        105.8$\pm$10.7 &                  19.6$\pm$8.6 &                 81.4$\pm$0.4 \\
          NGC  1667 &        80.5$\pm$2.7 &              5.3$\pm$3.5 &         175.3$\pm$8.9 &                   0.8$\pm$1.4 &        134.5$\pm$12.1 &                   0.2$\pm$0.7 &            270.5$\pm$4.3e-02 \\
          NGC  1672 &        86.2$\pm$3.7 &                1$\pm$0.6 &        328.1$\pm$23.0 &                     0$\pm$0.1 &          12.6$\pm$0.7 &                   0.2$\pm$0.2 &             53.3$\pm$1.1e-02 \\
          NGC  1808 &        50.9$\pm$1.2 &              2.9$\pm$1.9 &           nan$\pm$nan &           0.0e+00$\pm$0.0e+00 &          65.2$\pm$1.6 &                   3.3$\pm$3.7 &            147.4$\pm$1.2e-02 \\
          NGC  2110 &       100.2$\pm$2.3 &              4.1$\pm$2.1 &        372.5$\pm$15.2 &                   0.5$\pm$1.1 &         265.8$\pm$9.1 &                     2.1$\pm$2 &            210.0$\pm$1.2e-02 \\
           NGC  253 &        51.5$\pm$0.5 &              4.5$\pm$2.7 &         233.3$\pm$2.6 &                   0.1$\pm$0.1 &          42.5$\pm$0.5 &                   7.7$\pm$8.8 &            116.0$\pm$2.5e-04 \\
\enddata
\tablecomments{This table has been truncated for print. The full machine readable data table is available online. All estimates are derived from the pixels within the 200 pc nuclear aperture after data masking has been applied. Instancees of "--"  indicate no available data within the aperture after masking (see Section \ref{subsec:Data_Products}) was applied. Error values presented represent the standard error, calculated from the standard deviation of the pixels within the 200 pc nuclear aperture and scaled by the root of the number of pixels within the aperture.}
\end{deluxetable*}

\subsection{[Si VI] Elliptical Isophote Fitting}\label{subsec:A2}

Here (Table \ref{Table_A2}) we provide the data measuremnts for the [Si VI] isophote fitting, investigating the position angle, extent, and ellipticity of [Si VI] emission for each galaxy within the catalog. 

\begin{deluxetable*}{ccccccc}
\tabletypesize{\scriptsize}
\tablewidth{0pt} 
\tablecaption{[Si VI] Elliptical Isophote Fit Results.  \label{Table_A2}}
\tablehead{
\colhead{ } & \multicolumn{3}{c}{50$\%$ Max Intensity} &\multicolumn{3}{c}{10$\%$ Max Intensity} \\ 
\cline{2-7}
\colhead{Source Name} & \colhead{Ellipticity} &\colhead{Position Angle} &\colhead{SMA Length} & \colhead{Ellipticity} &\colhead{Position Angle} &\colhead{SMA Length} \\
\colhead{ } & \colhead{ } &\colhead{deg} &\colhead{pc} & \colhead{ } &\colhead{deg} &\colhead{pc}
} 
\colnumbers
\startdata 
Circinus        & 0.95 & 0.0    & 1.3   & 0.95 & 0.0   & 1.3   \\
ESO 137-G034    & 0.23 & -4   & 37.0  & 0.49 & -32 & 171 \\
ESO 428-G14     & 0.53 & -50  & 42.0  & 0.73 & -41 & 176 \\
ESO 548-81      & 0.12 & 17   & 21.3  & 0.13 & 90  & 85  \\
IC 2560         & 0.30 & 41   & 24.7  & 0.55 & 37  & 94  \\
IC 4296         & 0.32 & 25   & 384.7 & 0.11 & 46  & 535 \\
IC 4329a        & 0.22 & -79  & 16.9  & 0.24 & 47  & 79  \\
IC 5063        & 0.24 & -81  & 30.3  & 0.22 & -63 & 71  \\
IRAS 01475-0740 & 0.14 & -173 & 17.7  & 0.09 & 244 & 44  \\
MCG -05-23-016  & 0.11 & 48   & 21.0  & 0.14 & 32  & 63  \\
MCG -06-30-015 & 0.16 & 106  & 40.6  & 0.17 & 61  & 89  \\
MRK 1066        & 0.10 & -83  & 18.0  & 0.20 & 154 & 36  \\
MRK 1210        & 0.09 & -101 & 24.6  & 0.06 & 133 & 64  \\
MRK 573         & 0.39 & -150 & 66.6  & 0.20 & 146 & 109 \\
NGC 1068        & 0.66 & -13  & 24.4  & 0.35 & 6  & 44  \\
NGC 1194        & 0.10 & -36  & 13.4  & 0.23 & 10   & 47  \\
NGC 1320        & 0.06 & -165 & 13.4  & 0.18 & 94  & 45  \\
NGC 1365        & 0.11 & -51  & 15.2  & 0.09 & -46 & 50  \\
NGC 1386        & 0.49 & 6    & 29.5  & 0.36 & 3   & 59  \\
NGC 2110        & 0.32 & -32  & 21.6  & 0.54 & -20 & 181 \\
NGC 262         & 0.31 & -99  & 65.3  & 0.42 & 101 & 163 \\
NGC 2911        & 0.35 & 19   & 151.3 & 0.45 & 15  & 233 \\
NGC 2992        & 0.13 & 59   & 38.3  & 0.13 & 41  & 103 \\
NGC 3081        & 0.21 & -82  & 25.7  & 0.44 & -8  & 77  \\
NGC 3227        & 0.24 & 89   & 31.7  & 0.27 & 79  & 69  \\
NGC 3281        & 0.39 & -10   & 64.9  & 0.36 & 9   & 189 \\
NGC 3393        & 0.27 & -17  & 44.6  & 0.36 & 82  & 210 \\
NGC 3783        & 0.10 & -19  & 23.2  & 0.25 & 4   & 70  \\
NGC 4051        & 0.25 & -79  & 4.5   & 0.59 & 63  & 34  \\
NGC 4151        & 0.38 & 25   & 4.6   & 0.37 & 69  & 21  \\
NGC 4261        & 0.09 & 22   & 53.4  & 0.10 & 20  & 187 \\
NGC 4303        & 0.20 & 13   & 27.6  & 0.07 & 51  & 51  \\
NGC 4388        & 0.06 & -31  & 7.8   & 0.48 & 23  & 44  \\
NGC 4501        & 0.20 & 3    & 20.4  & 0.20 & 3   & 24  \\
NGC 4593        & 0.16 & -64  & 16.1  & 0.04 & -90 & 39  \\
NGC 4594        & 0.62 & -88  & 76.4  & 0.34 & -86 & 153 \\
NGC 4748        & 0.23 & -80  & 29.2  & 0.15 & -16 & 83  \\
NGC 4945        & 0.28 & 86   & 1.4   & 0.38 & 34  & 10   \\
NGC 5128        & 0.28 & 156  & 1.4   & 0.54 & -1  & 5   \\
NGC 513         & 0.28 & -8   & 29.3  & 0.21 & 69  & 205 \\
NGC 5135        & 0.14 & 57   & 92.3  & 0.17 & 42  & 221 \\
NGC 5506        & 0.16 & 73   & 17.3  & 0.34 & 52  & 61  \\
NGC 5643        & 0.28 & 83   & 25.6  & 0.36 & 81  & 56  \\
NGC 5728        & 0.24 & -75  & 26.0  & 0.36 & -55 & 85  \\
NGC 591         & 0.52 & -108 & 54.3  & 0.20 & -56 & 124 \\
NGC 6300        & 0.17 & 79   & 20.6  & 0.16 & 75  & 37  \\
NGC 676         & -- & --  & --   & 0.51 & 65.5  & 26  \\
NGC 6814        & 0.30 & -41  & 19.3  & 0.23 & -34 & 53  \\
NGC 6967        & 0.10 & -59  & 30.9  & 0.22 & 66  & 97  \\
NGC 7130        & 0.15 & 79   & 96.4  & 0.13 & 85  & 231 \\
NGC 7172        & 0.18 & 76   & 16.4  & 0.24 & 51  & 74  \\
NGC 7213        & 0.00 & -90  & 0.0   & 0.10 & 64  & 32  \\
NGC 7410        & 0.00 & -90  & 0.0   & 0.31 & 42  & 205 \\
NGC 7469        & 0.27 & -70  & 25.8  & 0.14 & -16 & 72  \\
NGC 7496        & 0.07 & -62  & 10.9  & 0.11 & 19  & 25  \\
NGC 7582        & 0.23 & -45  & 17.7  & 0.14 & -62 & 51  \\
NGC 7682        & 0.78 & -95  & 77.2  & 0.73 & 44  & 197 \\
NGC 7743        & 0.20 & 88   & 46.5  & 0.04 & 36  & 128 
\enddata
\tablecomments{Measurements in which a fit could not be achieved are indicated by "--". a: Upper limit, value represents 3$\sigma$ error level.}
\end{deluxetable*}

\subsection{Azimuthal Average – Seyfert Type}\label{subsec:A3}

\begin{figure*}[ht]
\plotone{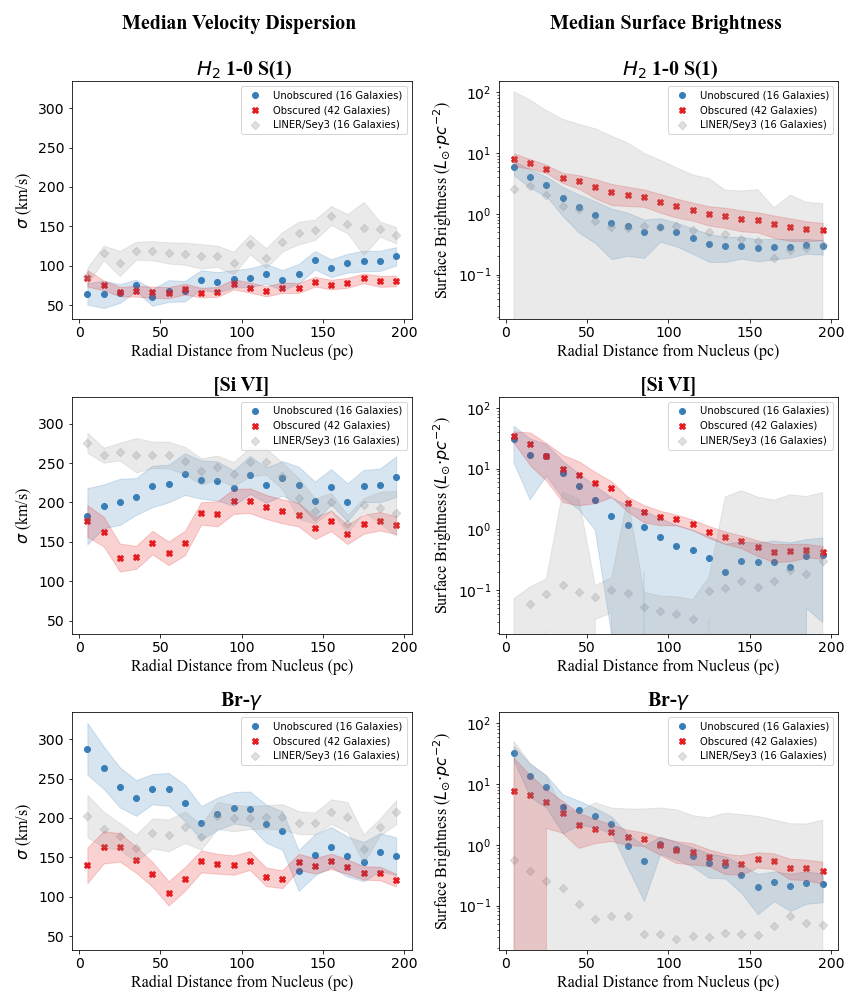}
\caption{Azimuthal average of velocity dispersion (left panel) and surface brightness (right panel) for each Seyfert type data.  Each point represents the median value of the mean data distribution at radial distance from the AGN.  The representative error bar presented depicts the average standard deviation of the across the sample set. 
\label{figA1}}
\end{figure*}

We include here plots of the the Azimuthal average for both surface brightness and $\sigma$ for the data binned by obscuration (Figure \ref{figA1}). 

\subsection{Azimuthal Average – X-ray Luminosity }\label{subsec:A4}

\begin{figure*}[ht]
\plotone{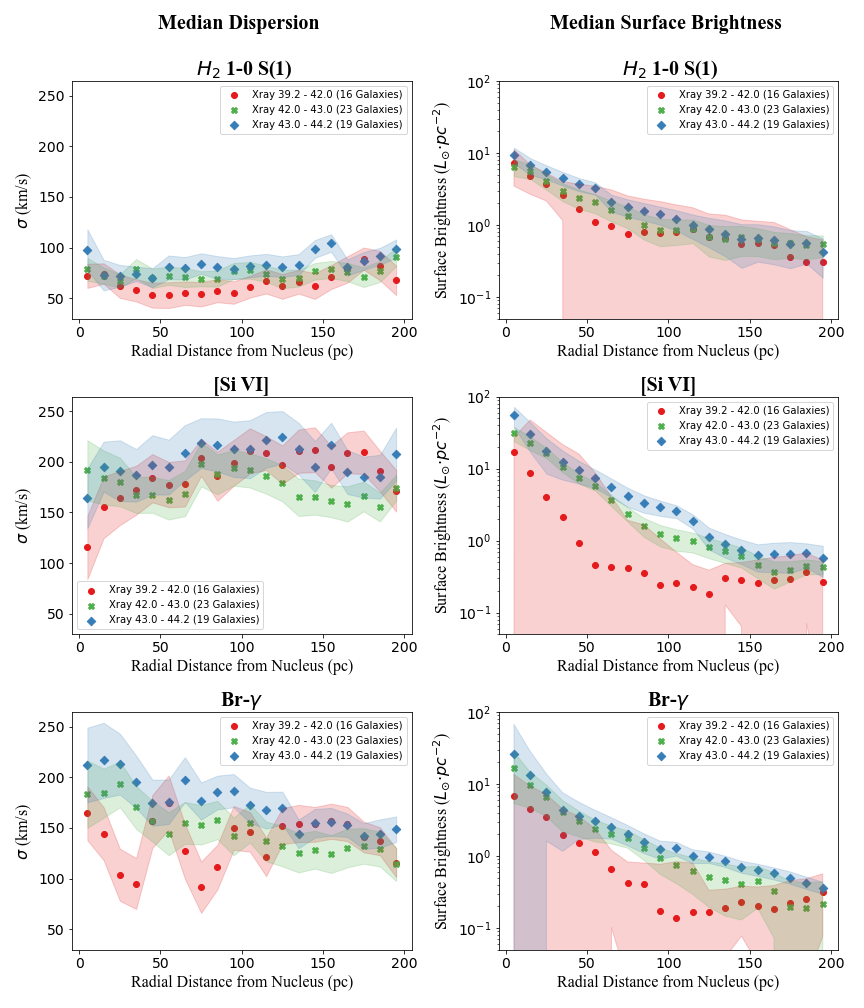}
\caption{Azimuthal average of velocity dispersion (left panel) and surface brightness (right panel) for each Seyfert type data.  Each point represents the median value of the mean data distribution at radial distance from the AGN.  The representative error bar presented depicts the average standard deviation of the across the sample set. 
\label{figA3}}
\end{figure*}

We include here plots of the the Azimuthal average for both surface brightness and $\sigma$ for the data binned by X-ray luminosity (Figure \ref{figA3}). 

\bibliographystyle{aasjournal}



\end{document}